\documentclass[epj]{svjour}
\usepackage{graphics}
\usepackage{color}

\newcommand{\be}{\begin{equation}}
\newcommand{\ee}{\end{equation}}
\newcommand{\bea}{\begin{eqnarray}}
\newcommand{\eea}{\end{eqnarray}}
\newcommand{\fc}{f_{\rm c}}

\newcommand{\leddth}{L_{\rm Edd}}

\newcommand{\kappae}{\kappa_{\rm e}}
\newcommand{\kappaR}{\kappa_{\rm R}}

\newcommand{\rmd}{{\rm d}}

\newcommand{\me}{m_{\rm e}}
\newcommand{\Ne}{N_{\rm e}}
\newcommand{\sigmat}{\sigma_{\rm T}}

\def\etal{{et\thinspace al.\,}}

 \newcommand{\apjs}{ApJS }  
 \newcommand{\apj}{ApJ }  
 \newcommand{\apjl}{ApJ }  
 \newcommand{\aap}{A\&A }  
  \newcommand{\aaps}{A\&AS }  
 \newcommand{\mnras}{MNRAS }  
 \newcommand{\pasj}{PASJ } 
  \newcommand{\nat}{Nat }  
   \newcommand{\physrep}{Phys. Rep. }  
    \newcommand{\jqsrt }{JQSRT }  
     \newcommand{\sovast}{Sov. Astr. }  
     \newcommand{\actaa}{Acta Astr. }

\begin{document}
%
\title{Measuring the basic parameters of neutron stars using  model
  atmospheres}  
\author{V.~F.~Suleimanov\inst{1,2} \and J.~Poutanen\inst{3,4} \and D.~Klochkov\inst{1} \and K.~Werner\inst{1}
}                     
%
%
\institute{Institut f\"ur Astronomie und Astrophysik, Kepler Center
  for Astro and Particle Physics,
Universit\"at T\"ubingen, Sand 1,
 72076 T\"ubingen, Germany
 \and
 Kazan Federal University, Kremlevskaja str., 18, Kazan 420008, Russia
\and
Tuorla Observatory, Department of Physics and Astronomy, University of Turku, V\"ais\"al\"antie 20, FI-21500 Piikki\"o, Finland
\and 
Nordita, KTH Royal Institute of Technology and Stockholm University, Roslagstullsbacken 23, SE-10691 Stockholm, Sweden
  }
\date{Received: date / Revised version: date}
%
\abstract{Model spectra of neutron star atmospheres are nowadays 
  widely used to fit the observed thermal X-ray spectra of  neutron
  stars. This fitting is the key element in the method of the neutron
  star radius  determination. Here, we present the basic assumptions used for the
  neutron star atmosphere modeling as well as the main qualitative
  features of the stellar atmospheres leading to the 
deviations of the emergent model spectrum from blackbody. We describe
the properties of two of our model atmosphere grids: (i) pure
carbon atmospheres for relatively cool neutron stars 
(1--4 MK) and (ii) hot atmospheres with Compton scattering
taken into account. The results obtained by applying these  grids to model 
the X-ray spectra of the central compact object in supernova remnant HESS\,1731$-$347, and two X-ray bursting neutron stars in
low-mass X-ray binaries, 4U\,1724$-$307 and 4U\,1608$-$52, are presented.
Possible systematic uncertainties
associated with the obtained neutron star radii
are discussed. 
%
} 
\titlerunning{Neutron star atmosphere models}
\maketitle
\section{Introduction}
\label{intro}

The fundamental problem driving the study of neutron stars
(NSs) is associated with unknown physical properties of super-dense
matter in the cores of these objects. It is also known as the 
problem of the neutron star equation of state (EOS). To determine the
EOS, one would need to measure simultaneously the mass and radius of
a NS. Existing theoretical models of EOS predict different 
mass-radius relations for NSs \cite{HPY07}. Various 
observational methods have been developed to constrain this relation using
observations \cite{LP07,2013RPPh...76a6901O}. 

One of the well developed methods is based
on fitting the  NS X-ray spectral continua 
with numerically computed model NS atmosphere spectra. 
Here, we present 
the basic ideas of this method and the
fundamental properties of NS atmospheres important for accurate
measurements of the NS masses and radii. We consider atmospheres of weakly
magnetized NSs only (surface magnetic field $B < 10^8 - 10^9$\,G).

Many atmosphere models have been computed almost twenty years ago,
shortly after the first isolated NSs had been discovered  \cite{1996Natur.379..233W}. 
 Most of the groups \cite{ZPS96,1996ApJ...461..327R,2009ASSL..357..181Z} 
 have computed NS atmospheres composed of light elements (hydrogen and
 helium) because heavy elements should have
sunk due to gravitational separation 
\cite{1980ApJ...235..534A,1983A&A...121..259H,2002ApJ...574..920B}. 
A number of works is dedicated to fitting the thermal X-ray spectra of isolated
NSs and the NSs in low-mass X-ray binaries (LMXBs) in 
quiescence situated in  globular clusters with known distances 
to restrict their masses and radii  
 \cite{HR06,2013ApJ...772....7G,2013ApJ...764..145C}.

The same model NS spectra were fitted to the observed X-ray spectra of the
point-like soft X-ray sources in the centers of some supernova 
remnants -- a special group of NSs referred to as central
compact objects (CCOs) \cite{1998A&A...331..821Z}.  The X-ray spectrum
of the youngest NS of this type located in Cas\,A 
was also fitted with hydrogen model atmospheres \cite{2009ApJ...703..910P}.
These authors found that this 
model assuming a homogeneously emitting stellar surface
gives a size of the emitting region of
about 4 -- 5.5\,km, compatible with a strange quark star only. 
A possible solution to this problem was suggested by Ho \& Heinke \cite{2009Natur.462...71H}, 
who have found 
that a pure carbon NS atmosphere model gives 
a reasonable NS size of 10 -- 14\,km. Moreover, the same authors later
have claimed an unusually 
rapid cooling of the CCO in Cas\,A \cite{2010ApJ...719L.167H,2011MNRAS.412L.108S,2013ApJ...777...22E}, 
but  this result has been questioned by Posselt \etal \cite{2013ApJ...779..186P}.
Recently,  model spectra of pure carbon NS atmospheres were
successfully applied to fit 
the X-ray spectrum of the CCO in \mbox{HESS\,J1731$-$347} \cite{2013A&A...556A..41K,2015A&A...573A..53K} using our models 
\cite{2014ApJS..210...13S}. We describe the obtained results in
the current review. 

X-ray bursting NSs in LMXBs are also widely used to measure the NS
masses and radii  \cite{LvPT93}. 
Thermonuclear burning  at the bottom of the freshly accreted matter
during an X-ray burst can be  
so intense that the luminosity $L$ reaches the Eddington limit $L_{\rm Edd}$. 
This can be used to obtain additional constraints on the NS mass and radius
\cite{Ebi87,1987A&A...172L..20V,1987MNRAS.226...39S,1989MNRAS.236..545F,Damen90,vP90}.


Strong photon-electron interaction in hot NS atmospheres of X-ray bursting NSs leads to blackbody-like emergent spectra  
\cite{London86,1986SvAL...12..383L,1991SvA....35..499Z,1991MNRAS.253..193P,Madej91,SPW11,SPW12}, 
such that the observed spectra of X-ray bursts can be fitted well with a blackbody \cite{GMH08}.  
In fact, the emergent spectra are not blackbody  but
are close to the diluted blackbody. 
Recently, we computed extended grids of hot NS atmospheres in a
wide range of the input parameters \cite{SPW11,SPW12,2015A&A...581A..83N}.
Based on these computations, we also proposed a new cooling tail
method for determination of the NS 
mass and radius using X-ray bursts with a photospheric radius
expansion \cite{SPW11,SPRW11}.
This method is described  in our review in detail as well as the
results of the application of this method 
to the NSs in LMXBs 4U\,1724$-$307  \cite{SPRW11} 
and 4U\,1608$-$52 \cite{2014MNRAS.442.3777P}. 
For the most recent analysis of a carefully selected sample of bursts and the 
resulting constraints on the equation of state, we refer to ref. \cite{2015arXiv150906561N}. 

 In principle, the atmosphere model spectra can be directly fitted to the data 
to obtain the constraints on the NS parameters 
\cite{1989A&A...220..117K,2005AcA....55..349M},
but the quality of the data as well as a weak dependence of the spectra on gravity 
will preclude that in the nearest future.

\section{Radius measurements of an ideal neutron star}
\subsection{Spectral fitting}

 Let us consider an isolated non-rotating thermally emitting
homogeneous neutron star with an intrinsic luminosity $L$, 
mass $M$ and radius $R$ at a known distance $D$.
In this case, the spectral flux density $F_{\nu}$ at  Earth can be computed as
\be \label{norm} 
    F_\nu = {\cal{F}}_{\nu,\infty} \, K = {\cal{F}}_{\nu,\infty} \frac{R_\infty^2}{D^2} = \frac{{\cal{F}}_{\nu(1+z)}}{1+z} \frac{R^2}{D^2}.
\ee 
Here $K=(R_\infty/D)^2$ is the normalization factor and 
\be \label{spinf}
{\cal{F}}_{\nu,\infty} = \frac{{\cal{F}}_{\nu(1+z)}}{(1+z)^{3}}
\ee
is the undiluted observed spectral flux density, formally attributed to the
apparent neutron star surface. The connection (\ref{spinf}) between ${\cal{F}}_{\nu,\infty}$
and the  intrinsic NS emergent  
spectral flux density ${\cal{F}}_\nu$ (the spectrum) can be derived using Liouville's theorem for photons 
(see, e.g. \cite{RL79,1983bhwd.book.....S}).
The \emph{apparent} NS radius measured by an observer at infinity
$R_\infty =R(1+z)$ is larger than the 
intrinsic radius because of light bending, and  the gravitational
redshift $z$ is related to the NS parameters as 
\begin{equation} \label{eq:redshift_def}
    1+z=(1-2GM/c^2R)^{-1/2} .
\end{equation}

The intrinsic spectrum $\cal{F}_{\nu}$ can be computed theoretically, and 
in generally depends on the effective temperature $T_{\rm eff}$, NS gravity $g$ and chemical composition. 
The effective temperature is related to 
the intrinsic bolometric flux ${\cal{F}}=\int {\cal{F}}_{\nu} \rmd \nu$ as: 
\be
 \sigma_{\rm SB} T_{\rm eff}^4 = {\cal{F}} = \frac{L}{4\pi R^2},
\ee
where $L$ is the NS luminosity measured at the surface
and $\sigma_{\rm SB}$ is Stefan-Boltzmann constant. 
The surface gravity accounting for the general relativity effect is 
\begin{equation} \label{eq:g_def}
   g=\frac{GM}{R^2}(1+z) .
\end{equation} 

The bolometric observed flux can be easily obtained from eq.\,(\ref{norm}): 
\begin{equation} \label{eq:fbol_bb}
F= \int F_{\nu} \rmd \nu =    \frac{R_\infty^2}{D^2} \frac{ {\cal{F}}}{(1+z)^4} = K \frac{ \sigma_{\rm SB} T_{\rm eff}^4}{(1+z)^4}.
\end{equation}
Thus if we know the observed flux as well as the redshifted, intrinsic NS spectrum ${\cal{F}_{\nu,\infty}}$ (or the 
redshifted effective temperature $T_{\rm eff}/(1+z)$), and the distance $D$, we can find the apparent NS radius $R_\infty$. 
A curve of a constant apparent radius in the $M - R$ plane 
(shown  by the dashed line in fig.\,\ref{sv_f1}) can constrain
the NS radius, because the mass is limited within a
relatively narrow range 1.2$-$2 $M_\odot$. 
It is clear, however, that the shape of the emitted spectrum is equally 
important for this method as the distance to the NS.

\subsection{Eddington luminosity}

\begin{figure}
\resizebox{0.5\textwidth}{!}{%
  \includegraphics{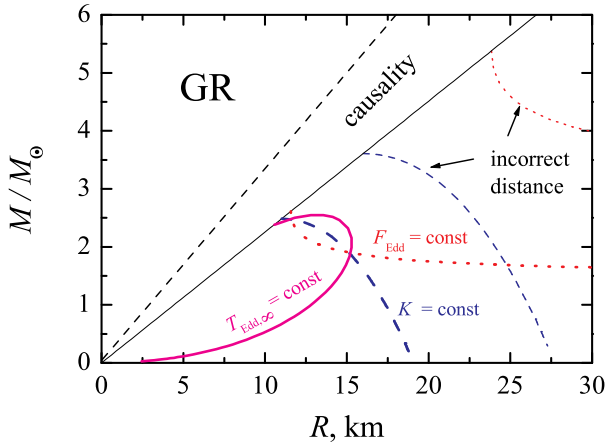}
}
\caption{The constraints on $M$ and $R$ of the neutron star from various
  observables. The solid curve gives the relation obtained for the
  Eddington temperature $T_{\rm Edd,\infty} = 1.64\times 10^7$\,K  
using eq.\,(\ref{tedd}) for a pure hydrogen atmosphere ($X = 1$). 
The dotted curves are for the Eddington flux $F_{\rm
  Edd}=5.61\times 10^{-8}$\,erg\,s$^{-1}$\,cm$^{-2}$ (using
eq.~(\ref{eddfl})),  
dashed curves are for $K$=1.37$\times 10^{-32}$ (from
eq.~(\ref{norm})), shown  for  
two different distances, $D$ = 5.3\,kpc (thick curves) and 7.7 kpc
(thin curves). For a larger distance there are no solutions 
(the curves shown by thin lines do not intersect).}
\label{sv_f1}   
\end{figure}

We consider NSs that can reach the Eddington luminosity, 
which is determined by the balance of the surface gravity $g$
and the radiative acceleration at the NS surface
\begin{equation} \label{eq:grad_def}
   g_{\rm rad} = \frac{1}{c} \int_0^\infty \kappa_\nu\,{\cal{F}}_{\nu,\rm Edd}\,\rmd\nu,
\end{equation}
where $\kappa_\nu$ is the opacity of the NS atmosphere. Because the
opacity depends on the local physical parameters of the matter, 
such as temperature $T$, density $\rho$, and chemical
composition, the Eddington luminosity is 
specific to every radiating astrophysical object. 
Therefore, the coherent Thomson electron scattering opacity  
\begin{equation} \label{eq:kappae}
\kappae = \sigmat \frac{\Ne}{\rho} \approx 0.2\ (1+X) \ \mbox{cm}^2\ \mbox{g}^{-1}  
\end{equation}
is commonly used to determine  the Eddington luminosity.  Here
$\sigmat= 6.65\times 10^{-25}$ cm$^2$ is the Thomson cross-section,  $\Ne$  is the electron number
density, and $X$ is the hydrogen mass fraction. 
The Thomson opacity does not depend on the local matter
properties because $\Ne \sim \rho$, but depends on $X$. 
It allows us to define the intrinsic bolometric
Eddington flux  
\be    
   {\cal{F}}_{\rm Edd} = \int_0^\infty {\cal{F}}_{\nu, \rm Edd}\  \rmd \nu = \frac{gc}{\kappae},
\ee 
and the intrinsic Eddington luminosity 
\begin{equation} \label{eq:ledd_def}
   \leddth= 4\pi R^2 \,{\cal{F}}_{\rm Edd} =\frac{4\pi GMc}{\kappae} (1+z). 
\end{equation}
The observed Eddington luminosity is lower:
\be
    L_{\rm Edd, \infty} = \frac{L_{\rm Edd}}{(1+z)^2} = \frac{4\pi GMc}{\kappae} (1+z)^{-1}, 
\ee
here one  factor $1+z$ accounts for the photon energy redshift, while another for the time dilation. 
The observed  bolometric Eddington flux at Earth can also be defined as
\be   \label{eddfl}
    F_{\rm Edd} = \frac{L_{\rm Edd, \infty}}{4\pi D^2}.
\ee
Therefore, if we know that the NS luminosity is equal to the Eddington
luminosity at a certain time and measure the bolometric flux 
at this time, we obtain additional constraints on the NS mass and
radius. Apart from the distance, the hydrogen mass fraction $X$ is
another crucial uncertainty here. The corresponding curve 
$L_{\rm Edd, \infty}=const$ in the $M-R$ plane is also shown by the dotted curve in
fig.\,\ref{sv_f1}. 
Because both  $R_\infty$ and $L_{\rm Edd, \infty}$ depend on the distance $D$,
the corresponding curves  $R_\infty = const$ and $L_{\rm Edd, \infty}= const$ are also distance dependent. 
The correct values of $M$ and $R$ of the NS are located
at the crossing points of the two curves.

It is easy to combine two observed values, $F_{\rm Edd}$ and $K$,
which both depend on distance $D$, to one value, which is
independent of $D$  
\be \label{tedd}
  \frac{F_{\rm Edd}}{K}  = \frac{gc}{\kappae}(1+z)^{-4} = \frac{{\cal{F}}_{\rm Edd}}{(1+z)^4} = \sigma_{\rm SB} \left(\frac{T_{\rm Edd}}{1+z}\right)^4.
\ee
The new quantity $F_{\rm Edd}/K$ is simply the redshifted intrinsic
Eddington flux ${\cal{F}}_{\rm Edd, \infty} = {\cal{F}}_{\rm  Edd}(1+z)^{-4}$. 
Here, we also introduced the Eddington effective temperature
$T_{\rm Edd}$, which is a suitable parameter describing the Eddington flux.

The contour $T_{\rm Edd, \infty}=T_{\rm Edd}/(1+z) = const$  in the $M-R$ plane 
shown in fig.\,\ref{sv_f1} by the solid curve corresponds to all possible 
crossing points of the curves $K= const$ and $F_{\rm Edd}= const$ at
all possible values of $D$.  This curve allows to evaluate the NS
radius even without any information about the distance. 
We note, however, that the result does depend on the assumed hydrogen mass fraction which 
affects the Thomson opacity and thus the Eddington limit. 
 
\section{Neutron star model atmospheres}
\subsection{Basic properties}
\label{sect:basic}

The atmosphere of a neutron star is a relatively thin outer plasma
envelope, where the escaping radiation is formed. 
To obtain a correct model emergent spectrum, a complete physical model
of the atmosphere needs to be computed.  
Therefore, all the hydrodynamical conservation laws such as mass,
energy, and momentum conservation  (the Euler equations) have to be
taken into account together with an appropriate 
equation of state of the matter. 

\subsubsection{Steady state approximation}

Usually standard steady state
atmospheres without any large-scale movements are considered, and the
Euler equations are reduced  
to the hydrostatical equilibrium equation.  The envelope lies above a
source of energy, a hot crust of an isolated cooling NS
 or a thermonuclear burning layer in X-ray bursting NS, and does
 not have any own local energy sources. Therefore, the thermal energy is just 
 transported through the atmosphere under a local energy
 balance. Energy is transported by radiation and a  
 correct  radiative transfer description is the most important part of
 atmosphere modelling.   A possible energy transport by  
 convection or by electron conductivity is ignored. Convection is
 possible in the NS envelopes at early stages of X-ray bursts. The
 electron conductivity
could be significant at the deepest atmosphere layers only, where the
electron degeneracy becomes significant. Fortunately, it only happens 
in the atmosphere layers, which are much deeper than those where
escaped photons originate even for low-temperature 
atmospheres  \cite{2014ApJS..210...13S}, not affecting the 
emergent spectra.  
   Another important approximation is the absence 
of any external flux of  radiation or fast particles bombarding the
outer atmospheric boundary.   
  
 A detail description
 of the NS atmospheres modelling is given in the Sect. \ref{numdet}.
Here, we describe the structure of NS atmospheres
which determine the emergent model spectra.
     
\subsubsection{Plane-parallel approximation and hydrostatic equilibrium}

The thickness of NS atmospheres is negligible compared to the NS
radius. A typical pressure-scale height is about 
\be
         h_{\rm p}  \approx \frac{kT}{\mu\, m_{\rm H}\, (g-g_{\rm rad})} = 1.3\, T_6 \,(g-g_{\rm rad})_{14}^{-1}\,\,\rm{cm}, 
\ee
where $\mu=0.62$ is the mean molecular weight of a particle in a fully
ionized solar mix plasma, and $m_{\rm H} = 1.67 \times 10^{-24}$\,g is
the proton mass (here and later, we use the notation for
some variables in cgs units $Q_x = Q\times 10^{-x}$).
Therefore, almost all NS atmospheres can be treated in the
plane-parallel approximation where all the physical  
variables, such as density $\rho$ and temperature $T$, depend on the
vertical geometrical coordinate $s$ only. 
Using the column density $m$ 
\begin{equation}
\rmd m = -\rho \, \rmd s \, ,
\end{equation}
instead of $s$ is even more suitable in this case. We note, that
$m=0$ corresponds to the upper atmospheric boundary.

Under the described assumptions, the hydrostatic equilibrium equation
\be \label{hydr}
      \frac{\rmd P}{\rmd m} = g-g_{\rm rad}
\ee
has the integral
\be
      P  =  \rho \frac{kT}{\mu\, m_{\rm H}} \approx m\,(g-g_{\rm rad}).
\ee
 Here, $P$ is the gas pressure and the equation of state for an ideal gas
 is assumed. We discuss below the applicability of this approximation.

 \subsubsection{Chemical composition}
 
 The chemical composition of the NS atmosphere is the most uncertain
 model parameter.  Atmospheres of non-accreting NSs 
have to be chemically pure and consist of the lightest chemical
element of the envelope only because of  
 gravitational separation \cite{1980ApJ...235..534A,1983A&A...121..259H,2002ApJ...574..920B}. 
The reason for this is a short characteristic sedimentation time (a
diffusion time through a pressure-scale hight $h_{\rm p}$) 
of any heavy element with $A/Z \approx 0.5$ relative to hydrogen for
NS atmosphere 
\be
   t_{\rm sed} \approx 0.04 \frac{\rho^{1.3}}{g_{14}^2 T_6^{0.3}}\,\, {\rm sec}
\ee 
(see \cite{2014PhyU...57..735P} and references there). 

Atmospheres of accreting NSs have the same mixture of elements as the
donor star if the accretion rate  
is much faster than the process of gravitational separation. A typical
atmosphere thickness is about $m_{\rm atm}\approx 100$\,g\,cm$^{-2}$,
and  this surface density would be displaced by accreted matter for
$t_{\rm sed}\approx$\,0.04 sec at a mass accretion rate 
\bea
\dot M_{\rm cr} & =& 4\pi R^2\, \frac{m_{\rm atm}}{t_{\rm sed}} \approx 3\times 10^{16}\, R_6^2\,\, {\rm g/s} \\ \nonumber
 &&\approx 5\times 10^{-10}\,R_6^2\,\,{\rm M_\odot/yr}. 
\eea 
This mass accretion rate provides a luminosity
$L_{\rm cr} \approx 6\times 10^{36}$\,erg/s, or $\approx 0.03\, L_{\rm
  Edd}$ for a NS with $R$= 10 km and $M=1.5 M_\odot$.   
This luminosity corresponds to an effective temperature of $T_{\rm
  eff} \approx 10^7$\,K, which is much higher than the blackbody
temperatures of the thermally emitted NS  
in low-mass X-ray binaries in quiescence \cite{1999ApJ...514..945R}. Therefore, the mass
accretion rate in quiescence in such systems, if accretion takes place, is 
much lower than $\dot M_{\rm cr}$ and it disturbs the pure hydrogen
composition only slightly.  
 On the other hand, in X-ray bursters the luminosity is higher and therefore 
the atmosphere composition is expected to be similar to that of the accreting matter.

The lightest element can be hydrogen for NS accreting matter from the
normal donor star or helium in ultra-compact binary systems  with a
helium white dwarf as a donor star \cite{2012MNRAS.423.1556S},
or even C and O from stripped C/O white dwarfs  \cite{2001ApJ...560L..59J}. 

\subsubsection{Opacity}

The properties of a model atmosphere as well as features of an
emergent model spectrum are determined by 
the interaction between photons and atmosphere particles. This
interaction is described by the plasma opacity 
$\chi_\nu$ and the optical depth $\tau_\nu$ in the stellar model
atmosphere theory  \cite{M78} 
\be
    \rmd \tau_\nu = \chi_\nu\, \rmd m.
\ee
The total opacity $\chi_\nu$ is divided in two physically different
parts, the electron scattering $\kappae$  and 
a ``true'' absorption opacity $\kappa_\nu$
\be
   \chi_\nu = \kappae + \kappa_\nu.
\ee   

The absorption opacity  $\kappa_\nu$ describes processes with photon
destruction, such as  photoionization (bound-free processes),  
transitions between electron energy levels in atoms and ions
(bound-bound processes), and interactions between free electrons 
and charged nuclei (free-free processes):
\be
     \kappa_\nu \sim \sigma_\nu \frac{\Ne N^+}{\rho}.
\ee
Here, $\sigma_\nu$ is the cross-section of a specific process and
$N^+$ is the  number density of the ions involved. 

The photon energy transforms to the thermal motion of plasma
particles in these processes. 
In the opposite ``true''  emission processes a new-born photon takes part
of the thermal energy of the particles and carries the information 
about the temperature in the birth-place into other parts of the
atmosphere. The absorption opacity is almost linearly proportional to
the local plasma density because  two charged particles participate in
these processes. 

A photon does not disappear in a scattering off an electron. It simply
changes its  propagation direction. The momentum and energy exchange between
the photon and  
electron are significant only for temperatures in excess of $10^7$\,K and photon
energies  $E=h\nu$ above 10 keV. Therefore, the electron scattering
in relatively cool NS atmospheres with $T_{\rm eff} \sim 10^6$\,K can
be considered as a coherent process and $\kappae$ does not depend on
the local properties in a fully ionized plasma. However, the energy
exchange in electron scatterings is much more significant in
luminous NS atmospheres with 
$T_{\rm eff} \ge 10^7$\,K, and determines the emergent spectrum. This
case of Compton scattering will be considered separately.  

\subsubsection{Plasma equation of state}

The equation of state of the ideal gas as well as the local thermodynamical
equilibrium (LTE) provide relatively good approximations for NS
atmospheres \cite{RSW08}.   
The LTE  assumption allows us to compute number densities of different
ionization states for all chemical species together 
with number densities of all exited (and ground) states of every
ionization state, using Saha and Boltzmann equations. Formally, this assumption
means that any transitions between different
levels 
of a given ionization state of a certain chemical element together with
the ionization and recombination processes 
are determined by interactions with free electrons only. This assumes
that the radiative transitions like photoionization and
photo-re\-com\-bi\-na\-tion are negligible compared to collisions with
electrons. In fact, the radiative transitions cannot be ignored, but the deviations
from  LTE are not significant because of the thermal-like radiation field in the NS atmospheres. 
Therefore,  LTE is a sufficiently good approximation for low-tem\-pe\-ra\-ture  
NS atmospheres because of their high densities and thermal radiation field. 
Almost all of the most abundant chemical elements are fully ionized
in high-tem\-pe\-ra\-ture NS atmospheres making the LTE assumption
reasonable. We can expect significant deviations from LTE in  
atmospheres of luminous NSs with enhancement abundances of iron group
chemical elements.   
The pressure ionization and level dissolution effects are important
for population computations, but they can be easily taken into account 
  using the occupation probability formalism \cite{1988ApJ...331..794H,1994A&A...282..151H},
  see also \cite{1994A&AT....4..307Z}. 

\subsubsection{Radiation field}

The radiation field in a given direction in model atmospheres is
described by the specific intensity $I_\nu \sim \nu^3\,N_\nu$, which
is connected with the photon occupation number $N_\nu$. The angular averaged
specific intensity is referred to as mean intensity $J_\nu$, and the first
momentum of $I_\nu$  
with respect to the angle $H_\nu$ describes the net rate of radiant
energy flow in the atmosphere 
\be
       J_\nu = \frac{1}{2} \int_{-1}^{+1} \,I_\nu\,\rmd \mu,~~~~~    H_\nu = \frac{1}{2} \int_{-1}^{+1} \mu\,I_\nu\,\rmd \mu,
 \ee  
where $\mu=\cos\theta$, and $\theta$ is the angle between photon
propagation direction and the normal to the plane-parallel
atmosphere. 
We note that the Eddington flux $H_\nu$ is connected to the physical
flux ${\cal{F}}_\nu$ as $4\pi H_\nu = {\cal{F}}_\nu$. 

The radiation transfer equation describes the interaction between
photons and plasma:
\be
    \mu\frac{\rmd I_\nu}{\rmd m} = (\kappa_\nu+\kappae)I_\nu - (\kappa_\nu\,B_\nu + \kappae\,J_\nu),
\ee
where $B_\nu$ is the Planck function. The first term on the right-hand
side of the equation describes the absorption  rate of photons,  
the second one -- the emission rate of photons in a given direction 
due to  ``true''  emission processes and coherent electron scattering.

\subsubsection{Thermal structure}

A stellar atmosphere can be divided in optically thin ($\tau_\nu <
1$) surface layers and optically thick deep layers ($\tau_\nu \gg 1$). 
In the optically thick layers, the thermodynamical
equilibrium  between radiation and plasma holds, such that $J_\nu \approx
B_\nu$. 
The temperature of these layers is determined by diffusion of
radiation, which is the first moment of the radiation transfer
equation integrated over frequency
\be
    \frac{\rmd B}{\rmd m} = \frac{3}{4\pi}\kappa_{\rm R}\,{\cal{F}}, ~~~{\rm or}~~~\frac{4}{3\kappa_{\rm R}}\frac{\rmd T^4}{\rmd m} = T_{\rm eff}^4.    
\ee 
Here $\kappa_{\rm R}$ is the Rosseland mean opacity of the plasma, which
does not depend on the radiation field and can be computed 
in advance. In the first approximation, the Rosseland opacity can be
expressed as a sum of the electron scattering opacity and
Kramers opacity: 
\be
     \kappa_{\rm R} \approx \kappae + 5\times 10^{24}\,\rho T^{-3.5}~~~{\rm cm}^2\,{\rm g}^{-1}.
\ee 
This approximation is valid for a relatively hot ($T > 10^4$\,K) solar
mix plasma  \cite{FKR02}.
The numerical coefficient in the Kramers law is significantly
smaller for a fully ionized pure hydrogen plasma, namely $\approx
6.6\times 10^{22}$.

An approximate solution for the model atmosphere temperature structure
is given by
\be  \label{grey}
         T (\tau_{\rm R}) \approx T_{\rm eff}\,(\tau_{\rm R} + q)^{1/4}.
\ee
Therefore, for the optically thick atmospheric layers, the deeper the hotter.
The constant $q$ is determined by the physical conditions in the
optically thin atmospheric layers. 
If the atmosphere absorption opacity $\kappa_\nu$ depends on photon
energies only slightly, the emergent spectrum will be close to 
the Planck function at the effective temperature, will form at
$\tau_{\rm R} \approx 1$ and $q \approx 0.5$. 
For example, this situation is realized in the Sun, where the
atmosphere opacity is determined by  
the negative hydrogen ion H$^{-}$, and depends on the photon energy in
the visual part of the spectrum only slightly.   

\subsubsection{Emergent spectra} 

There are two facts which qualitatively determine the emergent
spectrum. The emergent flux prefers escaping at the less opaque parts 
of the possible photon energy range. Numerically, it approximately
coincides with the Planck function at the averaged  
photon birth depth $m'_\nu$, ${\cal{F}}_\nu \approx \pi B_\nu
(T(m'_\nu))$. In typical NS atmospheres, the absorption opacity as well
as the photon birth depth strongly depend on photon
energy. Therefore, the birth depth could deviate significantly from 
the depth corresponding to $\tau_{\rm R} \approx 1$. 

\begin{figure}
\resizebox{0.5\textwidth}{!}{%
  \includegraphics{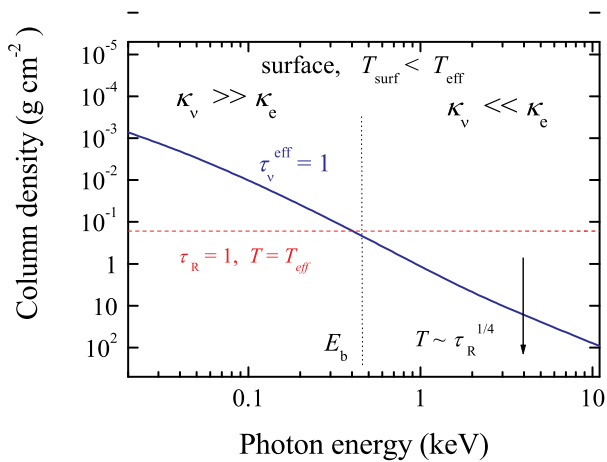}
}
\caption{The depth of the spectrum formation ($\tau_\nu^{\rm eff} =
  1$) vs. photon energy  
for a pure hydrogen atmosphere with $\log g$\,= 13.9 and $T_{\rm eff} = 1$\,MK.
The division to the low and high energy bands is also shown (see text). 
}
\label{sv_f2}   
\end{figure}

Let us consider a relatively cool pure hydrogen
NS atmosphere and qualitatively describe its emergent spectrum.
 Hydrogen in such an atmosphere is fully ionized as $E_{\rm ion}
 \approx 0.014$\,keV $\ll kT_{\rm eff} \approx 0.1$\,keV, 
and the absorption opacity is just the bremsstrahlung opacity
$\kappa_\nu \sim \rho\,\nu^{-3} \sim \rho\,E^{-3}$. 
It means that at every atmospheric depth $m$ there is a limiting
photon energy $E_{\rm b}$, such 
that $\kappa_\nu > \kappae$ at $E < E_{\rm b}$ and  $\kappa_\nu <
\kappae$ in the opposite case (see fig.~\ref{sv_f2}). 
The boundary energy is shifted to larger energies with increasing $m$ 
because of the increasing density.

\begin{figure}
\resizebox{0.5\textwidth}{!}{%
  \includegraphics{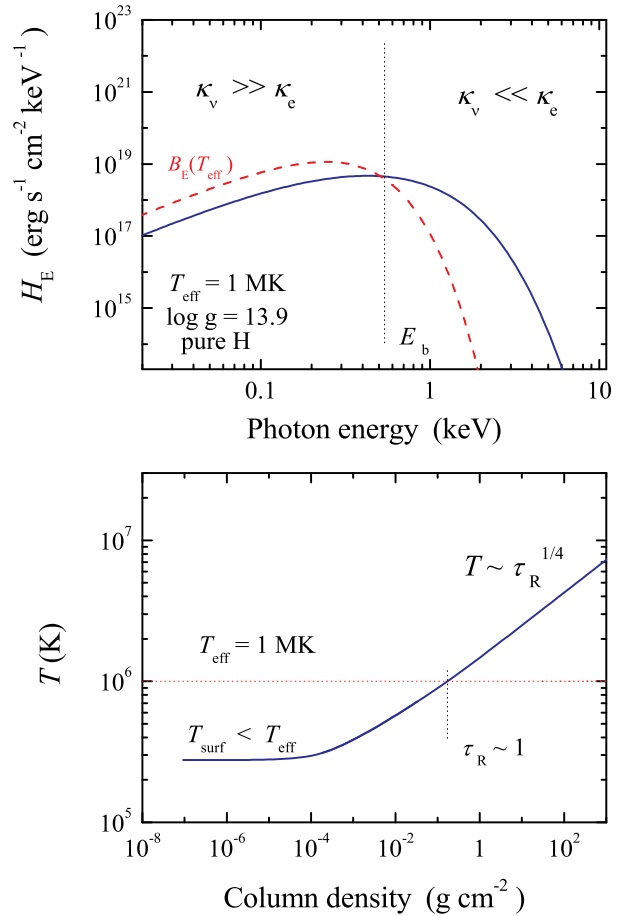}
}
\caption{The emergent spectrum ({\it top}) and temperature structure
  ({\it bottom}) of a pure hydrogen atmosphere with $\log g$\,= 13.9  
and $T_{\rm eff} = 1$\,MK. The corresponding blackbody spectrum
(dashed curve) is also shown in the top panel.}  
\label{sv_f3}   
\end{figure}
 
This allows us to divide the energy band in the atmosphere at
$\tau_{\rm R} \approx 1$ in such two 
parts. The escaping flux in the low energy part approximately equals
the Planck function at $\tau_\nu^a \sim \kappa_\nu\,m \approx 1$, 
${\cal{F}}_\nu \approx \pi B_\nu(T_{\rm surf})$. This optical depth
corresponds to the surface atmosphere layers 
because of $\kappa_\nu > \kappa_{\rm R}$ at this energy band. The
emergent flux corresponds in this case to the Planck function computed
for the surface temperature. 

In the high energy part, photons scatter a few times after being created in
a ``true'' emission processes before escaping because $\kappa_\nu < 
\kappae$. 
Therefore, the emergent flux here corresponds to the Planck function
computed for the temperature at the photon birth depth. Electrons 
scatter photons in random directions, and we cannot just take the
birth-depth at $\kappa_\nu\,m \approx 1$. The birth-depth for escaping
photons is not so deep and can be found using the so called effective
or  thermalization optical depth $\tau_\nu^{\rm eff} \sim
\sqrt{\kappa_\nu(\kappa_\nu + \kappae)}\,m \approx 1$ \cite{RL79}.  In the
considered energy band, the thermalization depths are 
larger than the depth $\tau_{\rm R} \approx 1$. Therefore, the
emergent fluxes are higher than the blackbody flux computed for the
effective temperature, 
${\cal{F}}_\nu \approx w_{\nu}\,\pi B_\nu(T > T_{\rm eff})$, where
$w_\nu \approx \sqrt{\kappa_\nu/(\kappa_\nu+\kappae)}$ is the dilution
factor.  
We note that the thermalization depth increases with increasing photon
energy such that the final emergent spectrum  
is much harder than $\pi B_\nu(T_{\rm eff})$ in the high-energy part.

Since the bolometric emergent flux is  conserved, the emergent
flux in the low energy part has to be lower than $\pi
B_\nu(T_{\rm eff})$. 
This means that the surface temperature should be significantly smaller
than  the effective temperature, up to $T_{\rm surf} \approx 0.2T_{\rm
  eff}$. Indeed, the surface temperature is determined by  
the balance between  heating due to absorption of photons  and
cooling due to emission of photons 
\be
       \int_0^\infty \kappa_\nu\,J_\nu \,\rmd\nu =  \int_0^\infty \kappa_\nu\,B_\nu(T_{\rm surf}) \,\rmd \nu.
\ee
The radiation field $J_\nu$  in the high energy band does not change
at the surface, and we know that it is hard.  But these hard photons
are absorbed 
only slightly by the plasma ($\kappa_\nu \sim \nu^{-3}$). This
absorbed energy can be easily compensated by much more effective
emission 
at low photon energies ($\kappa_\nu$ is large) even at a
relatively low plasma temperature. 

Examples of the numerically computed temperature structures and
emergent spectra of hydrogen NS model atmospheres are shown in
fig.\,\ref{sv_f3} to
illustrate the temperature structures and emergent spectra properties
described above. The helium model spectra are slightly
different. However, all the features discussed above are correct for
helium as well, because helium is also fully ionized in such atmospheres.
The model spectra of light element NS atmospheres were computed
by different authors (for review see ref. \cite{2009ASSL..357..181Z}). 

\subsubsection{Partially ionized atmospheres}

\begin{figure}
\resizebox{0.5\textwidth}{!}{%
  \includegraphics{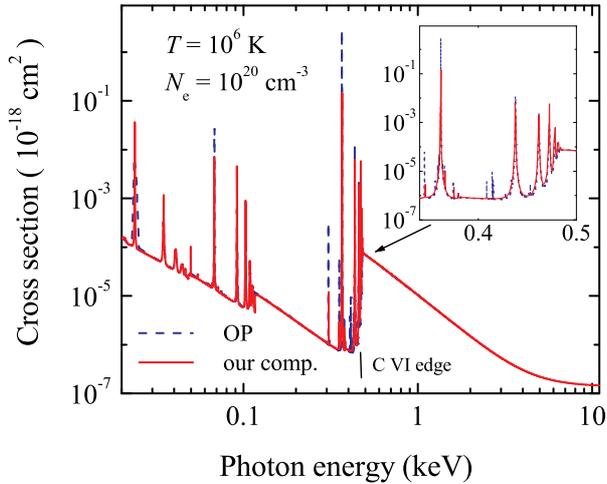}
}
\caption{Comparison of the carbon 
opacities published by \emph{Opacity Project} \cite{1994MNRAS.266..805S} (dashed curves) and
those computed by us (solid curves). 
The inset graph  shows a zoom into the region of the Lyman-like lines.
The  ionization threshold
energy of the C\,VI ion, 489 eV, is indicated
by the solid vertical line. From ref. \cite{2014ApJS..210...13S}}
\label{sv_f4}   
\end{figure}

\begin{figure}
\resizebox{0.5\textwidth}{!}{%
  \includegraphics{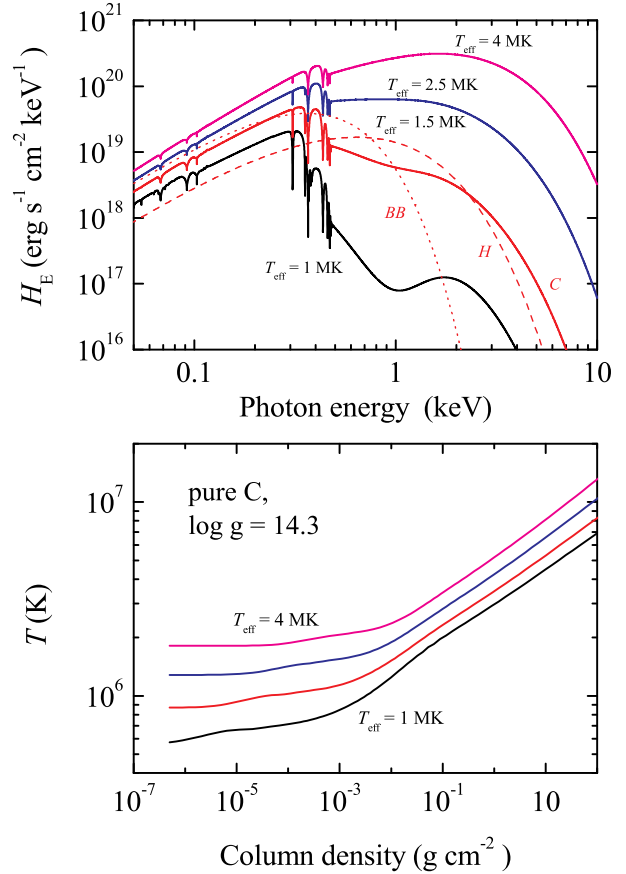}
}
\caption{Emergent spectra (top panel) and temperature structures 
(bottom panel) of pure carbon atmospheres with fixed $\log g$\,= 14.3  
and a set of effective temperatures. For $T_{\rm eff} = 1.5$\,MK, the
corresponding blackbody spectrum (dotted curve) and the pure 
hydrogen model spectrum (dashed curve) are also shown. From ref. \cite{2013ApJ...779..186P}}
\label{sv_f5}   
\end{figure}

Some NS envelopes might not have light elements, H or He.  Let us
consider a pure carbon atmosphere as an example. 
Carbon is not fully ionized in the considered cool NS atmospheres,
because the carbon ionization potential $E_{\rm ion}^{\rm C} \approx
0.5$\,keV is comparable to 
or higher than $kT_{\rm eff}$. Therefore, the absorption opacity is
much larger than in a fully ionized light element atmosphere because
photoionization cross-sections are larger than the bremsstrahlung
cross-section. Spectral lines also contribute significantly to the
total opacity (see fig.\,\ref{sv_f4}). 

The photoionization opacity is maximal at the photoionization edge and
decreases to higher photon energy. Therefore, in the total photon
energy band there is a relatively broad opaque band at $h\nu \ge
E_{\rm ion}^{\rm C}$. The emergent flux of a carbon atmosphere in this
opaque band forms in cool surface layers and carries a relatively
small fraction of the bolometric flux. As a result, photons  from a
carbon atmosphere escape in the more transparent higher 
 energy spectral band such that the high energy part of the spectrum
 is even harder than the corresponding part of the hydrogen/helium
 atmosphere spectrum. 
Such an ``opaque band'' effect is especially significant when the
maximum of the effective Planck function $B_\nu(T_{\rm eff})$ coincides
with the photoionization edge, $E_{\rm ion}^{\rm C} \sim 3kT_{\rm eff}$.

Examples of the temperature structures and the emergent spectra of carbon atmospheres are shown in fig.\,\ref{sv_f5}. 
By now two groups have computed extended grids of model spectra of NS carbon atmospheres \cite{2009Natur.462...71H,2014ApJS..210...13S}.

\subsubsection{Importance of Compton scattering in hot atmospheres} 

The approximation of coherent electron scattering breaks down for hot
luminous  NS atmospheres with  
$kT_{\rm eff} \ge 1$\,keV ($T_{\rm eff} \ge 10^7$\,K). Hence,
Compton electron scattering instead of  Thomson electron scattering 
has to be considered.

Let us again consider a hydrogen NS model atmosphere with $kT_{\rm
  eff} \approx 2$\,keV. The previously used division of the spectral
energy band  in two qualitatively different parts at the $\tau_{\rm R}
\approx 1$ depth -- 
the low energy band  with ``true''  absorption dominating over  electron scattering 
and the high energy band with  electron scattering as  main 
opacity source -- is still valid.

Photons created at $\tau_\nu^{\rm eff} \approx 1$ are still scattered
a few times before escaping. But now they exchange energy and momentum  
with cooler electrons in the surface atmospheric layers much more effectively. Therefore,  Compton
scattering supports thermodynamical equilibrium between photons and
electrons up to their escape. As a result, the hard part of the
emergent spectrum approaches a diluted blackbody spectrum  
\be
    {\cal{F}}_\nu \approx w\,\pi B_\nu(T_{\rm c})
\ee
with the colour temperature 
$T_{\rm c} > T_{\rm eff}$, because the averaged escaping depth has a
temperature higher than the effective temperature, and the number of
photons is not 
enough to form a normal Planck spectrum $B_\nu(T_{\rm c})$. Here $w
\approx \fc^{-4}$ is the dilution factor, and the colour correction
factor $\fc = T_{\rm c}/T_{\rm eff} > 1$.

\begin{figure}
\resizebox{0.5\textwidth}{!}{%
  \includegraphics{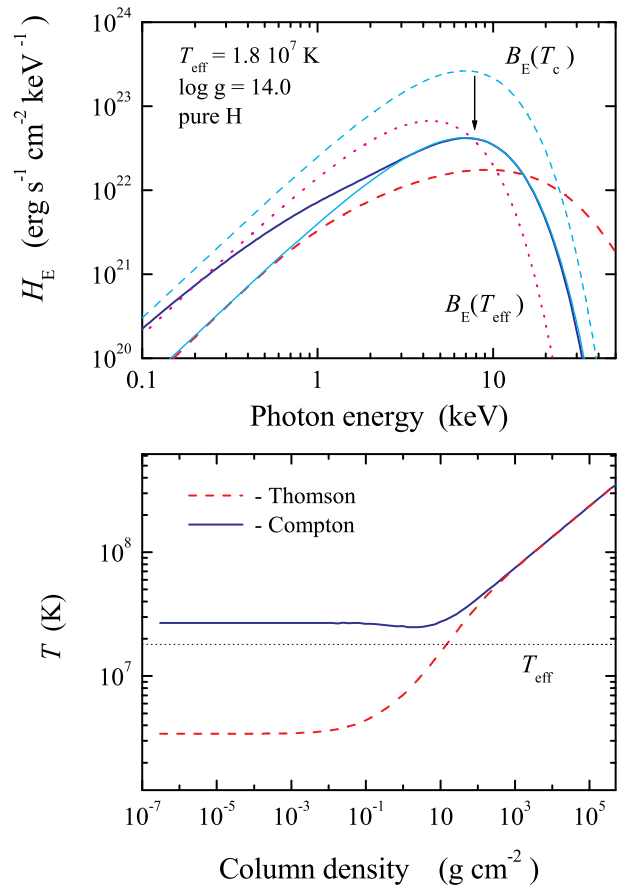}
}
\caption{The emergent spectra (top panel) and the temperature structures
  (bottom panel) of pure hydrogen atmospheres with  $\log g$\,=
  14.0 and $T_{\rm eff} = 18$\,MK computed taking  Compton
  scattering (black thick solid curves) into account, and using 
  Thomson scattering (red dashed curves). The corresponding blackbody
  spectrum (pink dotted curve), the diluted blackbody spectrum ($\fc=1.58$,
  blue thin solid curve), and the blackbody spectrum for the colour
  temperature (blue thin dashed curve) are also shown in the top panel. 
}
\label{sv_f6}   
\end{figure}

\begin{figure}
\resizebox{0.48\textwidth}{!}{%
  \includegraphics{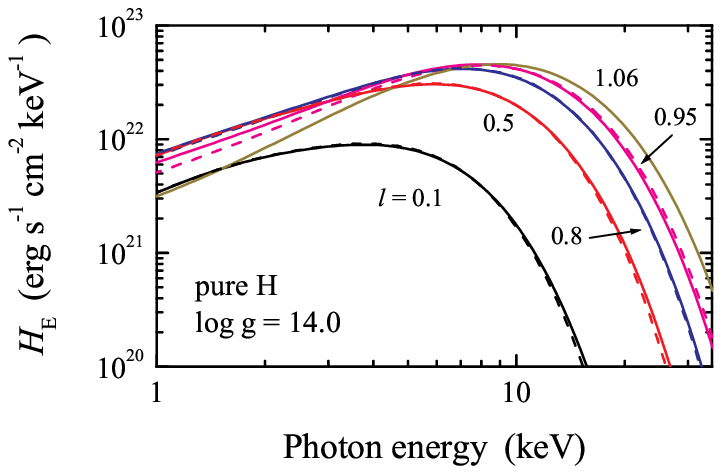}
  }
\resizebox{0.5\textwidth}{!}{%
  \includegraphics{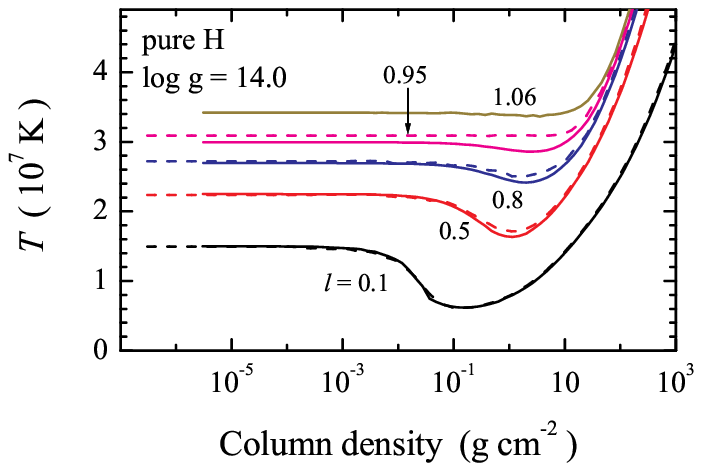}
}
\caption{The emergent spectra (top panel) and temperature structures
  (bottom panel) of pure hydrogen atmospheres with a fixed $\log g$\,=  14.0,  
and a set of relative luminosities. The models computed using  the
fully relativistic angular dependent redistribution function (solid
curves) are shown 
together with the models computed using Kompaneets equation (dashed curves).
From ref. \cite{SPW12}.
}
\label{sv_f7}   
\end{figure}

 Compton scattering also affects  the energy balance in the surface
layers. The approximate energy balance equation in the 
diffusion (Kompaneets) approximation is
\be
     \int_0^\infty \left(\kappa_\nu\,(J_\nu -B_\nu) -  \kappae \frac{4kT - h\nu}{m_{\rm e}c^2} J_\nu\right)\,\rmd \nu \approx 0.
   \ee
 The hard photons additionally heat the surface layer up to
$T_{\rm c}$ at the surface where the absorption opacity is negligible
because the density is low. Hence, the emergent spectrum in the low energy
band   
could be as high as $\pi B_{\nu}(T_{\rm eff})$ or even higher (see fig.\,\ref{sv_f6}). We note that
 Compton scattering  does not change the atmospheric temperature
structure in the deep optically thick layers, because  Compton
heating and cooling balance each other at $J_\nu \approx B_\nu$. 
Examples of the temperature structure and the emergent spectra of hydrogen
atmosphere models  with  Compton scattering taken into account 
for a set of relative luminosities $l=\cal{F}/{\cal{F}}_{\rm Edd}$
 are shown in fig.\,\ref{sv_f7}. 
 
\begin{figure}
\resizebox{0.5\textwidth}{!}{%
  \includegraphics{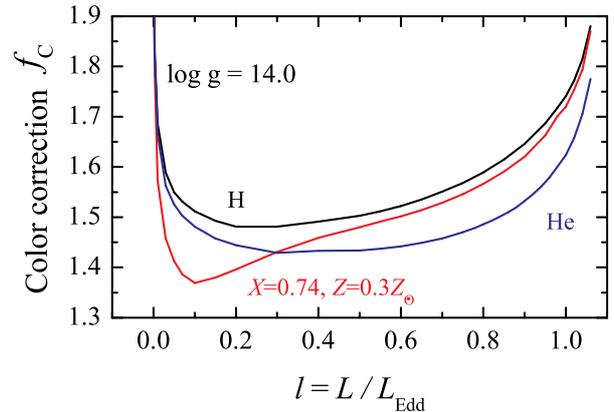}
}
\caption{The colour correction factors $\fc$ 
for the model atmospheres of three chemical compositions 
(pure hydrogen, pure helium, and solar hydrogen/helium mix with 30\%
of solar heavy-element abundance) 
as a function of the relative luminosity  $l$.  
}
\label{sv_f8}   
\end{figure}

 The colour correction factor $\fc$ is about 1.4$-$1.5 for moderately
 luminous NS atmospheres with   $l \sim 0.5$ and slightly depends  on
 the chemical composition of  the atmosphere (see fig.\,\ref{sv_f8}). But $\fc$ increases
 significantly, 
up to 1.7$-$1.9, for luminous atmospheres which are close to the
Eddington limit. The reason for that is a strong decrease of the
plasma density at $g_{\rm rad} \approx g$, see eq.~(\ref{hydr}). It
leads to a decrease of the absorption opacity and the thermalization
depth  is reached at relatively 
larger and hotter  column densities.

We have computed extended grids of model spectra of hot NS atmosphere for
several chemical composition and three values of $\log g$: 14.0, 14.3, and 14.6 
using two approaches: a diffusion approximation for Compton
scattering based on the Kompaneets equation \cite{SPW11} and 
using fully relativistic redistribution function to describe Compton scattering  \cite{SPW12}.   

\begin{figure}
\resizebox{0.5\textwidth}{!}{%
  \includegraphics{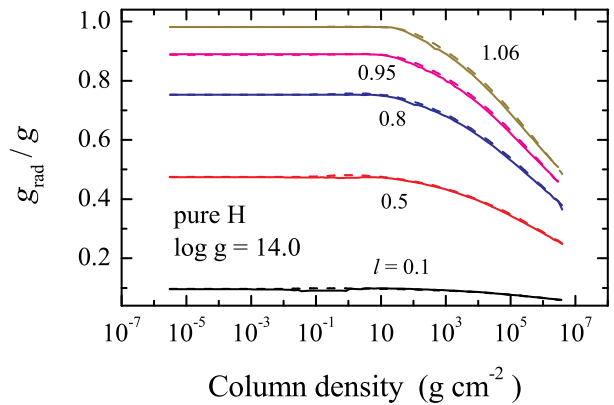}
}
\caption{Comparison of the relative radiative acceleration in pure
hydrogen atmospheres  computed exactly 
(solid curves) with the approximation given by eqs.~(\ref{eq:grad_lum}) and (\ref{eq:rossel}) 
and shown by the dashed curves (nearly coinciding with the solid curves). 
From ref. \cite{SPW12}.
} 
\label{sv_f9}   
\end{figure}

The model atmospheres in the latter work were computed for luminosities
formally exceeding the Eddington limit ($l > 1$) defined for Thomson opacity. The  reason 
is a well-known reduction of the electron scattering opacity at high temperatures (the Klein-Nishina effect). In fig.\,\ref{sv_f9}
we show the dependence of the ratio of the radiation force to the surface gravity $g_{\rm rad}/g$
on the depth of the atmospheric layers, which demonstrates clearly this
reduction (at small column densities $g_{\rm rad}/g$ is always smaller than $l$). 
We note that the ratios are smaller than unity for all 
relative luminosities, including $l > 1$.   

The radiative acceleration can formally be  represented as a product
of the flux- and  
temperature-dependent effective opacity
\be \label{eq:grad_opac}
g_{\rm rad} = \kappa (T)  \frac{\sigma_{\rm SB}T^4_{\rm eff}}{c}.
\ee
This expression can alternatively be written as 
\be \label{eq:grad_lum}
\frac{g_{\rm rad}}{g} = l\ \frac{\kappa (T)}{\kappae} . 
\ee
In the diffusion approximation, $\kappa (T)$ is given by the Rosseland
mean opacity, which is well fitted by 
the improved Paczynski's formula \cite{1983ApJ...267..315P,SPW12} 
\be 
\label{eq:rossel}
\kappaR (T) \approx \kappae \left[1+\left(\frac{kT}{39.4\ {\rm keV}} \right)^{0.976} \right]^{-1}  .
\ee

\begin{figure}
\resizebox{0.5\textwidth}{!}{%
  \includegraphics{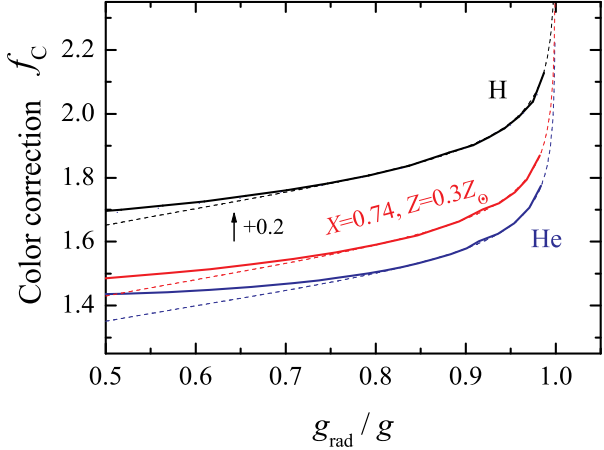}
}
\caption{The colour correction factor $\fc$ versus $g_{\rm rad}/g$ for the
  models of various chemical compositions.  
Solid curves are the results of calculations (for $\log g$ = 14.0), 
the dotted curves give the approximation (\ref{eq:fcfit}).  
For clarity, the curves for pure H models  are shifted upwards by 0.2. 
From ref. \cite{SPW12}.
}
\label{sv_f10}   
\end{figure}

The computed colour correction factors for luminous NS atmospheres are
well described by  the formula 
 \bea \label{eq:fcfit}
\fc &\approx& \left( \left[ 0.102+0.008X \right] \ln \frac{3+5X}{1-g_{\rm rad}/g} + 0.63-0.06X \right)
 ^{-4/5} \nonumber \\  
&\times& \left(  \frac{3+5X}{1-g_{\rm rad}/g} \right) ^{2/15} (g_{\rm rad}/g )^{3/20},
 \eea
which is similar to the formula suggested by Pavlov et al. \cite{1991MNRAS.253..193P} with 
numerical constants which also  depend  on the chemical composition.
This approximation works well for  $g_{\rm rad}/g > 0.8$ (see
fig.~\ref{sv_f10}).  

\subsection{Numerical details}
\label{numdet}

Under the assumptions described in Sect.\,\ref{sect:basic} and with the given
input parameters -- the effective temperature $T_{\rm eff}$, the
surface gravity $\log g$, and the chemical composition --
a model atmosphere has to be computed numerically. Here, we describe
some technical details for 
these computations. A more detailed description can be found in the papers
dealing with specific model atmospheres \cite{I03,SW07,SPW11,SPW12,2014ApJS..210...13S}. 
 
\subsubsection{Main equations} 

The structure of the NS atmosphere 
is described by a set of differential equations. The first one is the
hydrostatic equilibrium equation (\ref{hydr}).  
The second one is the radiation transfer equation for the
specific intensity $I(x,\mu)$ 
\begin{equation} \label{eq:rte}
\mu \frac{\rmd I(x,\mu)}{\rmd\tau (x,\mu)} = I(x,\mu) - S(x,\mu),
\end{equation}
where 
\begin{equation}
    \rmd\tau (x,\mu) = \left[\sigma(x,\mu)+k(x)\right]\, \rmd m ,
\end{equation}
$x=h\nu/m_{\rm e}c^2$ is the photon energy in units of electron rest mass,  and $k(x)$ is  the
``true'' absorption opacity. 
The source function is a sum of the thermal part and the scattering part
\be \label{eq:source}
S(x,\mu)  =   \frac{k(x) B_x +\kappae\,S_{\rm sc}(x,\mu)}{\sigma(x,\mu)+k(x)},   
\ee
where 
\begin{equation} \label{eq:planck}
B_x = B_{\nu} \frac{\rmd \nu}{\rmd x}, 
\end{equation}  
with $B_{\nu}$ being the Planck function. We note that all the intensities and source functions 
dependent on frequency are transferred  
using the same relation.
The scattering part of the source function $S_{\rm sc}(x,\mu)$ as well
as  the electron scattering opacity  
$\sigma(x,\mu)$ can be considered under three different assumptions.

In the simplest case of the coherent electron scattering, acceptable
at relatively low 
temperatures $T_{\rm eff} < 10^6$\,K, the electron scattering opacity
is given by eq.~(\ref{eq:kappae}), and 
the scattering part of the source function is just the mean intensity $J_x$ 
\be \label{cohsc}
    \sigma(x,\mu)=\kappae, \qquad S_{\rm sc}(x,\mu) = J_x.  
\ee

Compton scattering can be taken into account in the diffusion 
approximation using the Kompaneets equation 
\cite{SPW11} or using a fully relativistic
angle-depended redistribution function (RF) 
$R(x,\mu; x_{\rm 1},\mu_1)$, which describes the probability that a
photon with the dimensionless energy $x_1$  
propagating in the direction  corresponding to $\mu_1$ is scattered to
energy $x$ and in the
 direction corresponding to $\mu$ \cite{PS96,SPW12}. In our review, we consider the second 
 case only. The function $R(x,\mu; x_{\rm 1},\mu_1)$ is found by
 integrating over the azimuthal angle $\varphi$ of the RF
 $R(x,x_1,\eta$),  
which depends on the cosine of the angle between the 
directions of the photon propagation before and after scattering $\eta$
\bea \label{eq:rxxmu}
R(x,\mu; x_1,\mu_1) = \int_0^{2\pi} R(x,x_1,\eta)\,\rmd \varphi,\\ \nonumber 
\eta=\mu\mu_1+\sqrt{1-\mu^2}\sqrt{1-\mu_1^2}\cos\varphi.
\eea
The  RF depends on the depth $s$ via the electron temperature 
and satisfies the relation 
\begin{equation}
R(x_1,\mu_1; x,\mu) =  R(x,\mu; x_1,\mu_1)\, \exp\left(\frac{x-x_1}{\Theta}\right) ,
\end{equation}
which is the consequence of the detailed balance relation \cite{Pom73,NP94a}. 
Here,
\begin{equation} \label{ipc_u8}
 \Theta = \frac{kT}{\me c^2}
\end{equation}
is the dimensionless electron temperature.  

In this case, the electron scattering opacity is 
 \begin{equation}
\label{eq:scatopac}
  \sigma(x,\mu) \! = \!  \kappae \frac{1}{x} \!\int\limits_0^\infty\! x_1 \rmd x_1
 \!\!\!  \int\limits_{-1}^1 \!\!\rmd\mu_1 R(x_1,\mu_1;x,\mu)\,  
\! \left(1\!+\!\frac{C\,I(x_1,\mu_1)}{x_1^3}\right),
\end{equation}
where
\begin{equation} \label{eq:const_int}
C= \frac{1}{2m_{\rm e}} \left( \frac{h}{\me c^2} \right)^3,
\end{equation}
and the scattering part of the source function is 
\bea \label{eq:source2}
S_{\rm sc}(x,\mu)  & = &   \left(1+\frac{C\,I(x,\mu)}{x^3}\right)  x^2 \\
&  \times & \int_0^\infty
\frac{\rmd x_{\rm 1}}{x^2_{\rm 1}} \int_{-1}^1 \rmd\mu_{\rm 1} R(x,\mu;x_{\rm 1}, \mu_1) I(x_{\rm 1},
\mu_{\rm 1}). \nonumber
\eea

The formal solution of the radiation transfer equation (\ref{eq:rte})
is obtained using the short-characteristic method \cite{1987JQSRT..38..325O} in
three angles in each hemisphere.
The full solution is found with an accelerated $\Lambda$-iteration
method (see details in Appendix  
of ref. \cite{SPW12}).

The radiation pressure acceleration $g_{\rm rad}$ is computed using the RF as 
\bea \label{eq:grad}
&&g_{\rm rad} = \frac{\rmd P_{\rm rad}}{\rmd m} = 
\frac{2\pi}{c} \,\frac{\rmd}{\rmd m}\, \int^{\infty}_{0} \rmd x \, \int^{+1}_{-1}\mu^2 I(x,\mu)\, \rmd\mu \\ 
\nonumber
&&= \frac{2\pi}{c} \int^{\infty}_{0} \!\! \rmd x  \int^{+1}_{-1} \left[\sigma(x,\mu) + k(x)\right]  
\left[I(x,\mu)-S(x,\mu)\right]  \mu \rmd\mu, 
\eea
where the derivative with respect to $m$ is replaced by the first moment of the
 radiation transfer equation (\ref{eq:rte}).  
When the  source functions and the opacities are isotropic, this
expression is reduced to the standard definition  
\begin{equation} \label{eq:grad_stan}
g_{\rm rad} = \frac{4\pi}{c} \, 
\int^{\infty}_{0}   \left[\sigma(x) + k(x)\right]   H_x(m) \ \rmd x . 
\end{equation}
 
These equations are completed  by the energy balance equation
\begin{equation}  \label{eq:econs}
\int^{\infty}_{0} \! \! \rmd x  \! \! \int^{+1}_{-1} \! \! \left[\sigma(x,\mu) + k(x)\right]  \left[I(x,\mu)-S(x,\mu)\right] \rmd\mu = 0,
\end{equation}
the ideal gas law
\begin{equation}   \label{gstat}
    P = N_{\rm tot}\ kT,
\end{equation}
where $N_{\rm tot}$ is the number density of all particles, and  the
particle and charge conservation equations.   
The absorption opacity includes the free-free opacity 
as well as the bound-free transitions  
for all ions of the 15 most abundant chemical elements (H, He, C, N,
O, Ne, Na, Mg, Al, Si, S, Ar, Ca, Fe, Ni) \cite{I03}
using photoionization cross-sections from \cite{1995A&AS..109..125V}. Opacities due to spectral 
 lines can be considered for the cool NS models
with coherent electron scattering using $\sim 25\,000$ spectral lines from
the CHIANTI, Version 3.0, atomic database \cite{1997A&AS..125..149D}.

\subsubsection{Solution method} 

Our version of the computer code ATLAS \cite{K70,1993KurCD..13.....K}
modified to deal with high temperatures \cite{SulP06,SW07}, is
used to solve the above equations when the coherent electron
scattering is considered. 
The code was further developed to account for Compton scattering using
both the Kompaneets operator and the RF approach. 

We  can use various logarithmically equidistant frequency grids
describing the whole NS spectral band  
for a given $T_{\rm eff}$ in our computations. 
If the spectral lines are taken into account, an extended grid with
20\,000--40\,000 points are used. 
The methods considering Compton scattering does not permit
a large number of points, 
therefore,  grids  with 300--600  points  is used.
The grid of atmosphere depth layers consists of 98 depths $m_{\rm i}$ 
distributed equidistantly on a logarithmic scale from
$10^{-6}-10^{-8}$ to $m_{\rm max}=10^5-10^7$~g~cm$^{-2}$.  
The appropriate value of $m_{\rm max}$ is chosen to satisfy the inner
boundary condition 
of the radiation transfer problem $J_x \approx B_x$. The outer
boundary condition for the radiative transfer equation is 
the absence of any external radiation illuminating the atmosphere.

The first step of the calculations is the computation of an initial
grey atmosphere model (see eq.\,\ref{grey}) as well as  of the opacities at all 
depths and all frequencies. 
The solution of the radiative transfer equation (\ref{eq:rte}) was checked
for the energy balance equation (\ref{eq:econs}) together with the surface
flux condition
\begin{equation}
    4 \pi \int_0^{\infty} H_x ({m=0}) \rmd x = 4 \pi    H_0 = \sigma_{\rm SB} T_{\rm eff}^4 .
\end{equation}
The relative flux error 
\begin{equation}
     \varepsilon_{\rm H}(m) = 1 - \frac{H_0}{\int_0^{\infty} H_x (m) \rmd x},
\end{equation}
and the energy balance error 
\begin{equation}  \label{eq:econs1}
 \varepsilon_{\Lambda}(m) = \frac{1}{2}\! \int^{\infty}_{0} \!\! \! \rmd x \!\! \int^{+1}_{-1} \!\! 
\left[\sigma(x,\mu) + k(x)\right]  \left[I(x,\mu)-S(x,\mu)\right]  \rmd \mu
\end{equation}
 were calculated as  functions of depth. 
The temperature corrections were then evaluated using three different
procedures.  In the upper atmospheric layers, we used the integral
$\Lambda$-iteration method.  
The temperature correction for a particular depth was found as 
\begin{equation}
     \Delta T_{\Lambda} = -  \varepsilon_{\Lambda}(m)  \ 
     \left(\int_0^{\infty}
 \left[ \frac{\Lambda_{\rm d}(x)-1}{1-\alpha(x)\Lambda_{\rm d}(x)} \right]
k(x)\, \frac{\rmd B_x}{\rmd T}\, \rmd x  \right) ^{-1},
\end{equation}  
where $\alpha(x)=\sigma_{\rm CS}(x)/(k(x)+\sigma_{\rm CS}(x))$, and
$\Lambda_{{\rm d}}(x)$  is the diagonal matrix element of the $\Lambda$-operator. 
Here, $\sigma_{\rm CS}(x)$ is the Compton scattering opacity 
averaged over the relativistic Maxwellian electron distribution 
(see eq. (A16) in \cite{PS96}), which is  equivalent to
eq. (\ref{eq:scatopac})  
if one ignores the induced scattering), or just $\kappae$ for the
coherent electron scattering case. In the deep layers, we used the
Avrett-Krook flux correction based on the 
relative flux error $\varepsilon_{\rm H}(m)$. 
Finally, the third procedure was the surface correction based on the
emergent flux error 
(see ref. \cite{K70} for a detailed description of the methods).

The iteration procedure is repeated until the relative flux error gets
lower than 0.1\% and the relative flux  
derivative error gets smaller
than 0.01\%.   As a result,  we obtain a self-consistent NS  model
atmosphere, together with the emergent  spectrum of radiation. 
We note that this accuracy is unachievable for luminous models with
$g_{\rm rad} \approx g$. These models can have larger  relative flux
errors, up to 2--3\%.

\section{Applications to the real objects}

Below, we present the results of NS model atmospheres applications to the
real X-ray sources with NSs. We consider two classes of objects.
The fist class are thermally emitting  NSs in the centers of the
supernova remnants which do 
not show any evidence of accretion and of strong magnetic field. They
are referred to as central compact objects (CCOs). The second class is
comprised by X-ray bursting NSs in
low-mass X-ray binaries.

\subsection{Central compact objects in supernova remnants}

The first point-like soft X-ray sources in the centers of some supernova
remnants were assigned into a separate group of the
NSs (CCOs)  after the launch
of the {\sl Chandra} X-ray observatory \cite{2002ASPC..271..247P,2004IAUS..218..239P}. 
These sources  have thermal spectra 
($kT_{\rm BB} \sim 0.2 -0.6$\,keV) and relatively low luminosities ($L_{\rm x}
\sim 10^{33} -10^{34}$\,erg\,s$^{-1}$). They have not been detected in any other
electromagnetic wavelength range, from radio to $\gamma$-rays  \cite{2008AIPC..983..320G}.  Currently, eight
confirmed CCOs are known and three sources are considered to be CCO
candidates (see Table~1 of \cite{2013ApJ...765...58G}). 
Many of the CCOs have relatively good distance estimates, which is important for
measuring the NS radii  \cite{2002ASPC..271..247P,2004IAUS..218..239P,2013ApJ...765...58G}. 

The CCO in Cas~A is especially interesting. 
As mentioned in the Introduction, the carbon atmosphere has to be assumed to reconcile 
the derived size of the emitting region with that of a canonical NS \cite{2009Natur.462...71H}.
Recently, the X-ray spectrum of another CCO 
situated near the center of the supernova remnant shell HESS~J1731$-$347
(aka XMMU\,1732), 
was obtained with the observatory \emph{XMM-Newton} 
with a total exposure time $\sim 100$\, ksec. 
The source is  most probably located in the Scutum-Crux arm ($\sim$3\,kpc) with the
corresponding lower limit for the distance  of  $\sim$\,3.2\,kpc \cite{2011ApJ...735...12A}
derived on the basis of the X-ray absorption
pattern and the absorbing column densities based on CO observations. 
The source spectrum is also well described by the carbon
atmosphere model  \cite{2013A&A...556A..41K,2015A&A...573A..53K}.

We computed an extended grid of carbon NS model atmospheres to analyze
X-ray spectra of this CCO. 
 The models are computed for 9 values of surface gravity $\log g$, from
13.7 to 14.9 with a step of 0.15, which cover most of the realistic NS
equations-of-state for a wide range of NS masses. For every value of $\log g$,
we computed 61 models with $T_{\rm eff}$ from 1 to 4\,MK with a step of
0.05\,MK. This grid is available in form of an XSPEC model\\
\textsf{http://heasarc.gsfc.nasa.gov/xanadu/xspec/models/\\carbatm.html}.

\begin{figure}
\resizebox{0.45\textwidth}{!}{%
  \includegraphics{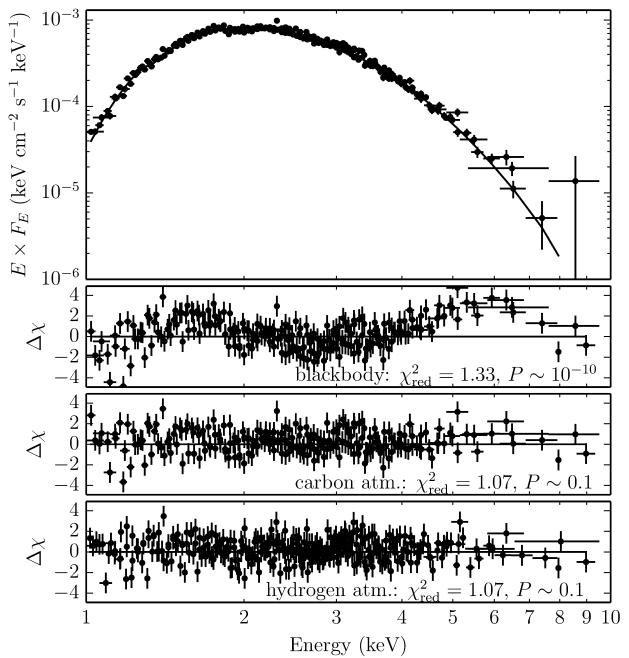}
}
\caption{ The combined \emph{XMM-Newton} spectrum (all three EPIC cameras) 
of the CCO in HESS~J1731$-$347  from the three available observations fitted with a carbon atmosphere model 
  (top panel) and the fit residuals for the blackbody, the carbon
  atmosphere, and the hydrogen atmosphere models (the lower three panels from top to
  bottom). The null-hypothesis probabilities ($P$-values) yielded by
  the $\chi^2$-fits are indicated. The blackbody model leads to a
  statistically unacceptable fit.
  From ref. \cite{2015A&A...573A..53K}.
  }
\label{sv_f11}   
\end{figure}

We assume that the carbon atmosphere covers
the entire NS located at a distance $D$ and used the direct fitting method
(see eq.\,\ref{norm}).  The NS mass  $M$ and radius $R$ are the
fitting parameters, and they 
can be easily translated to the gravitational redshift $z$ and the surface gravity $g$
using eqs.\,(\ref{eq:redshift_def}) and (\ref{eq:g_def}). 
During a fit, the model spectra are interpolated between the grid
values of $T_{\rm eff}$ and $\log g$.
The resulting 
model spectrum is compared to the observed one 
\be \label{norm2}
    F_{E} =  \frac{{\cal{F}}_{E(1+z)}(T_{\rm eff}, g)}{1+z} \, \frac{R^2}{D^2}.
\ee 
Here and later, we use the photon energy $E=h\nu$ instead of frequency
$\nu$ to describe the X-ray observations. 
The distance $D$ is formally a free parameter, but it can be fixed to obtain
a useful confidence region in the $M-R$ plane.   
 The model is multiplied by 
a component accounting for the interstellar photoelectric absorption
characterized by the equivalent column density of hydrogen atoms $N_{\rm H}$.

\begin{figure}
\resizebox{0.45\textwidth}{!}{%
  \includegraphics{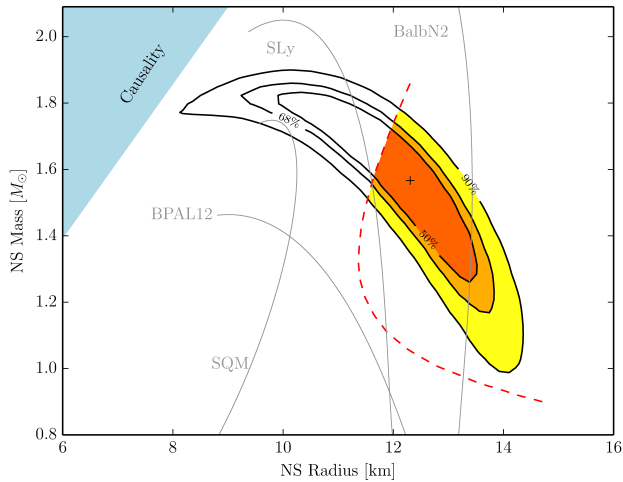}
}
\caption{The constraints on the mass and radius for the CCO in HESS~J1731$-$347 
assuming a distance of 3.2\,kpc. 
The region on the right hand side of the red dashed curve  is 
allowed by the cooling theory assuming a NS age of  27\,kyr (see text for details).
 From ref. \cite{2015A&A...573A..53K}.
   }
\label{sv_f12}   
\end{figure}

 We thus performed our spectral fitting for the
fixed distance of 3.2\,kpc using blackbody, pure carbon and hydrogen
model atmospheres. 
Both types of atmosphere models provide a
good fit to the data (fig.~\ref{sv_f11}). 
The best-fit parameters obtained for the carbon atmosphere models are
$M = 1.55^{+0.28}_{-0.24} M_\odot$, $R = 12.4^{+0.9}_{-2.2}$\,km,
$T_{\rm eff} = 2.24^{+0.39}_{-0.13}$\,MK, and $T_{\rm eff, \infty} = 1.78^{+0.04}_{-0.02}$\,MK.
The $\chi^2$ confidence regions in the $M-R$ plane (50, 68, and 90\%
levels) are shown in fig.\,\ref{sv_f12}. The hydrogen models give unreasonably
small $M$ and $R$.

\begin{figure}
\resizebox{0.45\textwidth}{!}{%
  \includegraphics{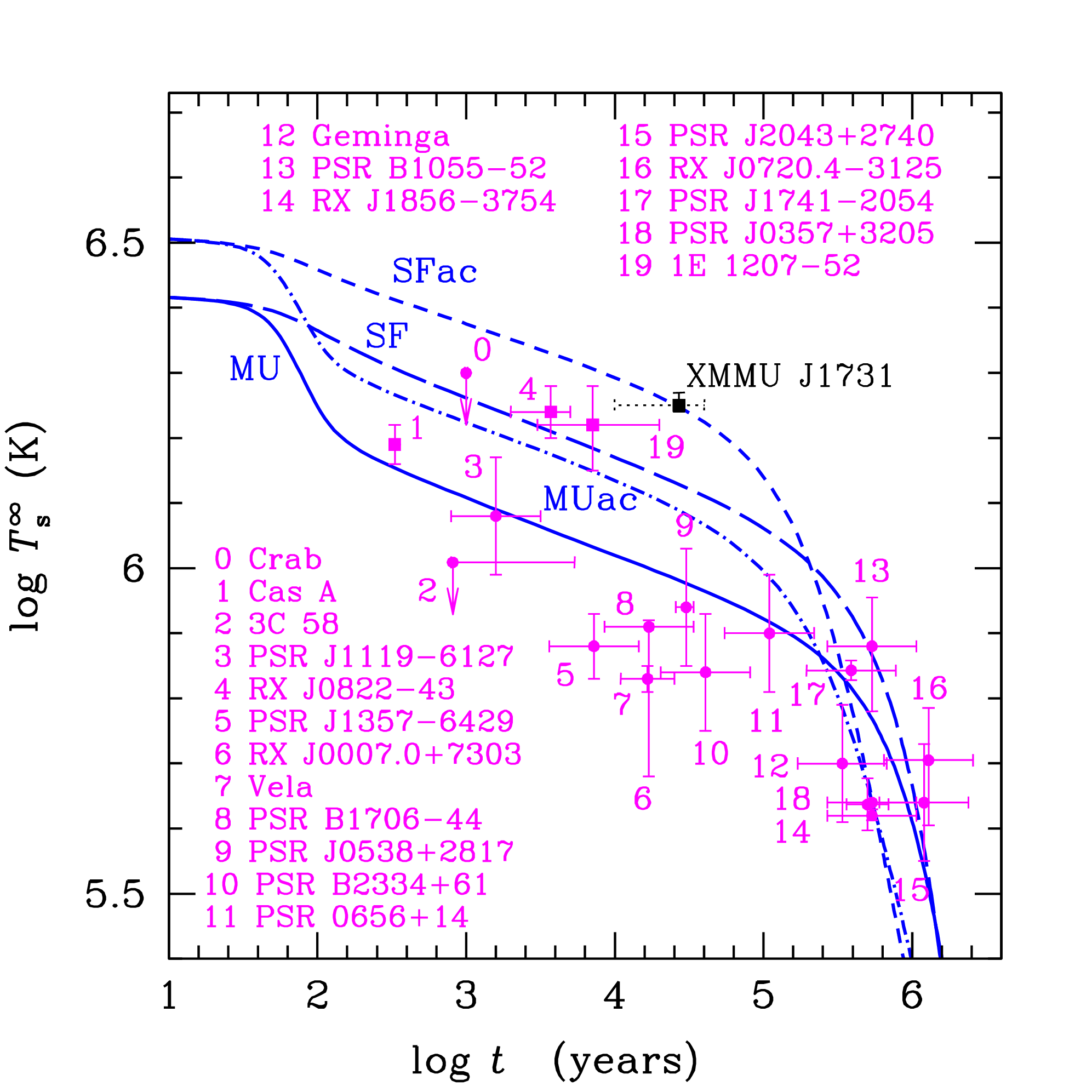}
}
\caption{ The effective surface temperatures or upper limits for a
     number of cooling
	isolated NSs including CCO in HESS~J1731$-$347 (marked as XMMU\,1731) versus their ages
        (data points) compared 
	with four theoretical cooling curves for a 1.5\,$M_\odot$ star.
        The plot contains four CCOs with reliable temperature
          measurements (labeled with squares). 
        For HESS~J1731$-$347, the dashed error bar indicates a conservative range
        of 10--40\,kyr adopted for the NS age.
	MU refers to a non-superfluid star with a heat
        blanket made of iron; 
	SF corresponds to a star with strong proton superfluidity in the core and
	the same heat blanket; MUac and SFac refer, respectively, 
	to the same models as MU and SF but with a fully carbon heat
        blanketing envelope (see text for details).
         From ref. \cite{2015A&A...573A..53K}.
                 }
\label{sv_f13}   
\end{figure}

 Unfortunately,  strong interstellar absorption \\ makes it difficult to directly observe 
carbon photoionization edges and the spectral lines, which are predicted by the atmosphere model. 
Potentially, their detection can give a possibility to determine the gravitational redshift, 
but long exposure times (up to a million second) might be needed for that. 
The problem might be even more severe if the magnetic field on the surface of the CCO in HESS~J1731$-$347 
is not as small as assumed, but above $10^{10} - 10^{11}$\,G. 
The magnetic field would shift the ionization energies as well as the energies of the spectral lines, 
making direct determination of the gravitational redshift impossible. 
Furthermore, presently there are no magnetized carbon model atmospheres available, and we cannot predict how
the magnetic field affects the fits.

The age of the CCO ($\sim 27$\,kyr) was estimated in ref. \cite{2008ApJ...679L..85T} 
assuming a distance of 3.2\,kpc. 
The uncertainties of the age estimate are not
well understood. We use a conservative range of
10--40\,kyr. For larger distances, the age must be larger. 
The CCO is very hot for this age compared  to other
isolated cooling NSs (see a $T_s^\infty-t$ diagram in fig.\,\ref{sv_f13}, where
$t$ is the estimated/measured age of the objects and $T_s^\infty
\equiv T_{\rm eff, \infty}$ is the apparent effective surface temperature).
This unusual property of the CCO in HESS~J1731$-$347 provides additional constraints on the
NS parameters 
using the current NS cooling theory (see  details and references in ref. \cite{2015A&A...573A..53K}).  
The CCO must have a very low neutrino luminosity   
and an unusually heat-transparent blanketing envelope for 
the observed slow cooling down.  The required low 
neutrino luminosity can be realized in a star with strong proton superfluidity
in the core (where neutrino emission is produced by neutron-neutron
collisions). High heat transparency can be provided by the presence of
a sufficient amount of carbon in the heat blanketing envelope. For a star's
age of 27\,kyr, the heat blanket should contain the maximum amount of carbon,
$\Delta M \sim 10^{-8}\,{\rm M_\odot}$.

The theoretical cooling curves  
depend (although not very strong) on $M$ and
$R$ for the observed 
$T_s^\infty$. Therefore, the observed $T_s^\infty$ and age cannot
be reconciled for certain $M$ and $R$, where the theoretical cooling
curve goes below 
the inferred $T_s^\infty$ even at the extreme parameters of the cooling
regulators. These values of $M$ and $R$ can be treated as
\emph{forbidden} by the cooling theory which gives
 additional constraints on $M$ and $R$. Under the assumptions
described in detail in ref. \cite{2015A&A...573A..53K},
the cooling theory mostly eliminates the 
range of small $R \le 12$\,km (see fig.\,\ref{sv_f12}).
 Recently, à more sophisticated cooling theory was used
for this kind of limitation  \cite{2015arXiv151000573O}, 
but this analysis almost did not change the basic conclusions.

\subsection{X-ray bursting neutron stars in low-mass X-ray binaries}

 X-ray bursting NSs are members of close binary systems with the
 secondary low-mass star overfilling its Roche lobe 
 (so called low-mass X-ray binaries, LMXBs) with a relatively low
 accretion rate, see reviews in ref. \cite{LvPT93,GMH08}. 
X-ray bursts are the observational manifestation of  thermonuclear
flashes  occurring  at the bottom of freshly accreted matter on
the NS surface. 
Some of them  can be  so powerful that the X-ray burst luminosity $L$
reaches the Eddington limit $L_{\rm Edd}$ leading
to the expansion of the photosphere  as measured by an increase of blackbody normalization $K$. 
Such photospheric radius expansion  (PRE)
bursts provide important information about  the NS
compactness from the observed Eddington flux and the maximum effective
temperature of the NS surface 
\cite{Ebi87,Damen90,vP90}. 

 The observed spectra of X-ray bursts are well-fitted by a  blackbody
  \cite{GMH08}
 \be
    F_{E} = \pi B_{E}(T_{\rm BB})\,K_{\rm BB} = \pi B_{E}(T_{\rm BB})\,\frac{R_{\rm BB}^2}{D^2}.
\ee 
In fact, the emergent spectra  are close to the diluted blackbody spectra 
because of strong photon-electron interaction in hot NS atmospheres, as it was described above
\be \label{eq:fit}
{\cal F}_{E} \approx \fc^{-4} \pi B_{E} (T_{\rm c} = \fc T_{\rm eff}).
\ee 
Recently, we suggested a new cooling tail method for NS mass and
radius determination using X-ray bursts with PRE \cite{SPW11,SPRW11}. 
This method is based on the fact that the observed 
normalization $K_{\rm BB}$ of an isolated passively cooling NS is proportional to
the dilution factor $\fc^{-4}$ only 
\be \label{kfc}
     K_{\rm BB} = \frac{R^2(1+z)^2}{\fc^4\,D^2}.
\ee
This conclusion can be easily obtained using this chain of reasoning
for the observed bolometric flux
\be
\sigma_{\rm SB}T_{\rm BB}^4\, K_{\rm BB} = \sigma_{\rm SB}T_{\rm eff, \infty}^4 \,\frac{R^2(1+z)^2}{D^2},
\ee
where the observed blackbody temperature can be expressed as
\be
       T_{\rm BB} = \frac{T_{\rm c}}{1+z} = \frac{\fc T_{\rm eff}}{1+z} = \fc T_{\rm eff, \infty}.
\ee
The approximation presented by eq.\,(\ref{eq:fit}) conserves bolometric flux
\be
      \mathcal{F} = \frac{1}{\fc^4}\sigma_{\rm SB}\,T_{\rm c}^4 = \sigma_{\rm SB}\,T_{\rm eff}^4,
\ee
therefore, we find the relation given by eq.\,(\ref{kfc}) between the observed normalization $K_{\rm BB}$ and the computed colour correction factor $\fc$.

\begin{figure}
\resizebox{0.45\textwidth}{!}{%
  \includegraphics{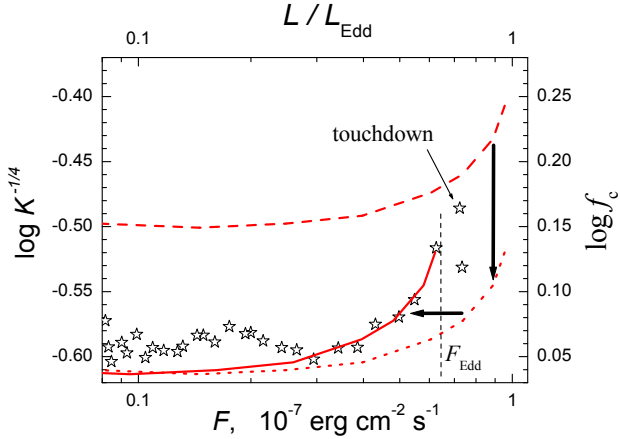}
}
\caption{Scheme of fitting of the observed dependence $K_{\rm BB}^{-1/4} - F_{\rm BB}$ (stars) by the computed dependence $\fc - l$ (dashed curve). 
  From ref. \cite{SPW11}.
    }
\label{sv_f14}   
\end{figure}

\begin{figure}
\resizebox{0.45\textwidth}{!}{%
  \includegraphics{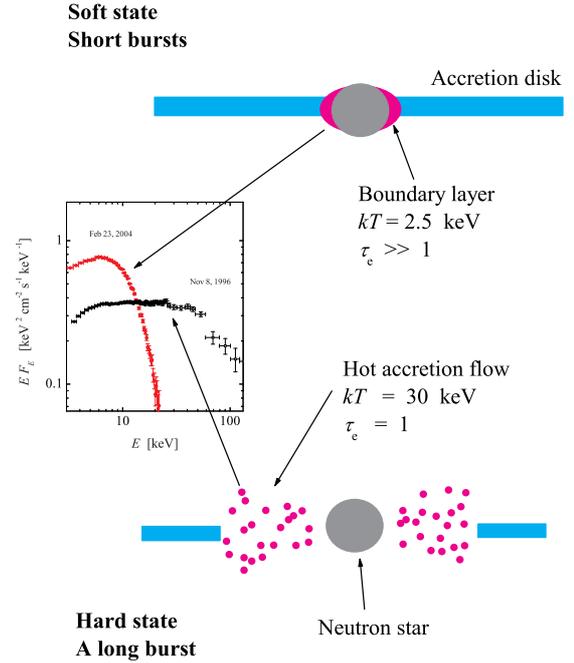}
}
\caption{A sketch of accretion structures around the NS in a LMXB in the
  soft and hard persistent states together 
with the corresponding persistent X-ray spectra before two PRE bursts in
4U\,1724$-$307 \cite{SPRW11}.   
  }
\label{sv_f15}   
\end{figure}

As mentioned above, the colour correction factor $\fc$ depends
mainly on the relative NS luminosity $l = L/L_{\rm Edd}$. 
Therefore, the observed normalization $K_{\rm BB}$ has to depend on the
observed bolometric flux $F_{\rm BB}$ in the same way as the
computed $\fc^{-4}$ depends on $l$. 
Therefore, we can fit the observed dependence $K_{\rm BB}^{-1/4} - F_{\rm BB}$
with the computed one $\fc - l$ in the cooling phase of the  
burst (see fig.\,\ref{sv_f14}) and obtain two fitting parameters
\be
 A =   \left(\frac{R(1+z)}{D}\right)^{-1/2},~~~~ {\rm so~that}~~~~K_{\rm BB}^{-1/4}= \fc A,
 \ee
and the observed Eddington flux
\be
   F_{\rm Edd} = \frac{L_{\rm Edd, \infty}}{4\pi D^2}=\frac{GM\,c}{\kappae \,D^2} \frac{1}{1+z}.
   \ee
They can be combined to the observed Eddington temperature which is independent of $D$
\be
  T_{\rm Edd, \infty} = A\left(\frac{F_{\rm Edd}}{\sigma_{\rm SB}}\right)^{1/4} = 9.81\,A'F_{\rm Edd, -7}^{1/4}\,{\rm keV},
\ee
where $F_{\rm Edd, -7} = F_{\rm Edd} / 10^{-7}$\,erg\,s$^{-1}$\,cm$^{-2}$, and 
the normalized fitting parameter $A'=(R_\infty{\rm [km]}/D_{\rm 10})^{-1/2}$ with
$D_{\rm 10} = D/ 10$\,kpc.
The Eddington temperature $T_{\rm Edd, \infty}$ found from observation  determines a specific
curve in the  $M-R$ plane (see fig.\,\ref{sv_f1}), which allows the NS radius to be
evaluated, as NS masses are in a  range 1.2 -- 2 $M_\odot$. 

\begin{figure*}
\begin{center}
\resizebox{0.85\textwidth}{!}{%
  \includegraphics{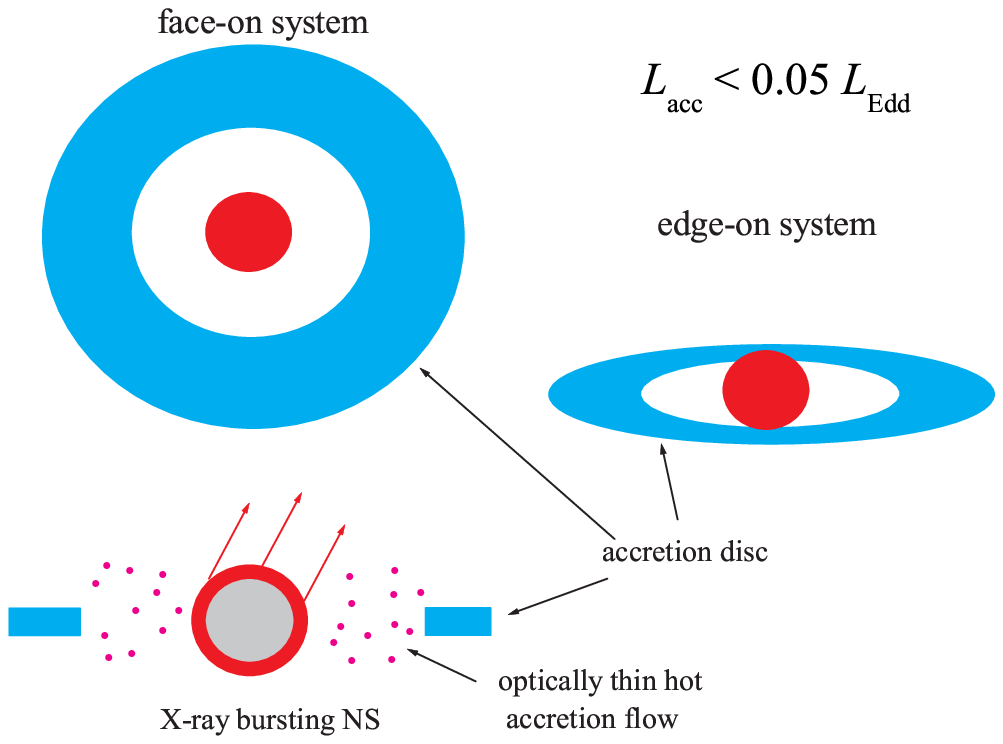}
 }
\includegraphics{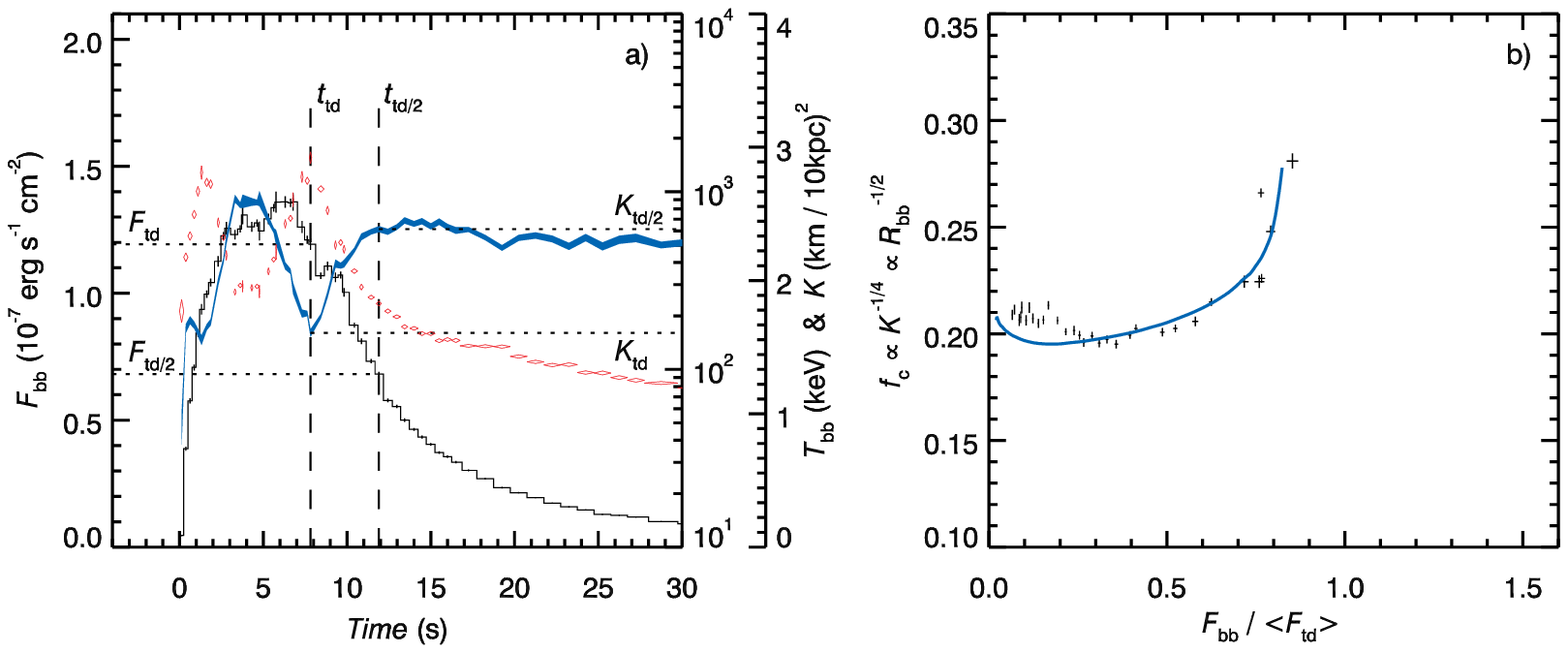}
\caption{Top panel: A sketch of the radial cross-section of
  accretion structures around the NS in the hard persistent state of LMXBs
together with the projections on the sky for low (face-on) and high
(edge-on) inclination systems. 
Bottom panel:  The time resolved spectroscopy of a PRE X-ray
burst from 4U 1608$-$52 
in the hard persistent state taken from ref. \cite{2014MNRAS.445.4218K}. 
In the left panel, the black line shows the bolometric flux $F_{\rm BB}$ 
(left-hand Y-axis).
The blue ribbon shows the $1\sigma$ limits of the black body normalisation 
$K_{\rm BB} = (R_{\rm BB} [{\rm km}] / D_{10})^2$ (inner right-hand y-axis). 
The red diamonds show the $1\sigma$ errors for black body temperature
$T_{\rm BB}$ in keV (outer right-hand y-axis).  
The first black vertical dashed line marks the time of touchdown $t_{\rm td}$
and the second vertical dashed line to the right shows the time
$t_{\rm td/2}$ when $F_{\rm BB}$ has decreased to one half of the touchdown
flux.
The corresponding $F_{\rm BB}$ and $K_{\rm BB}$-values at these times $F_{\rm td}$,
$F_{\rm td/2}$, $K_{\rm td}$ and $K_{\rm td/2}$ are marked with dotted lines.
The right panel shows the relationship between the inverse square root of
the black body radius (proportional to the colour-correction factor $\fc$) and
the black body flux $F_{\rm BB}$ that is scaled using the mean touchdown flux
$\langle F_{\rm td}\rangle$.
The blue line is the model prediction for a pure hydrogen NS atmosphere with a
surface gravity of $\log g = 14.3$ \cite{SPW12}.
 }
\label{sv_f16}  
\end{center} 
\end{figure*}

\begin{figure*}
\begin{center}
\resizebox{0.85\textwidth}{!}{%
  \includegraphics{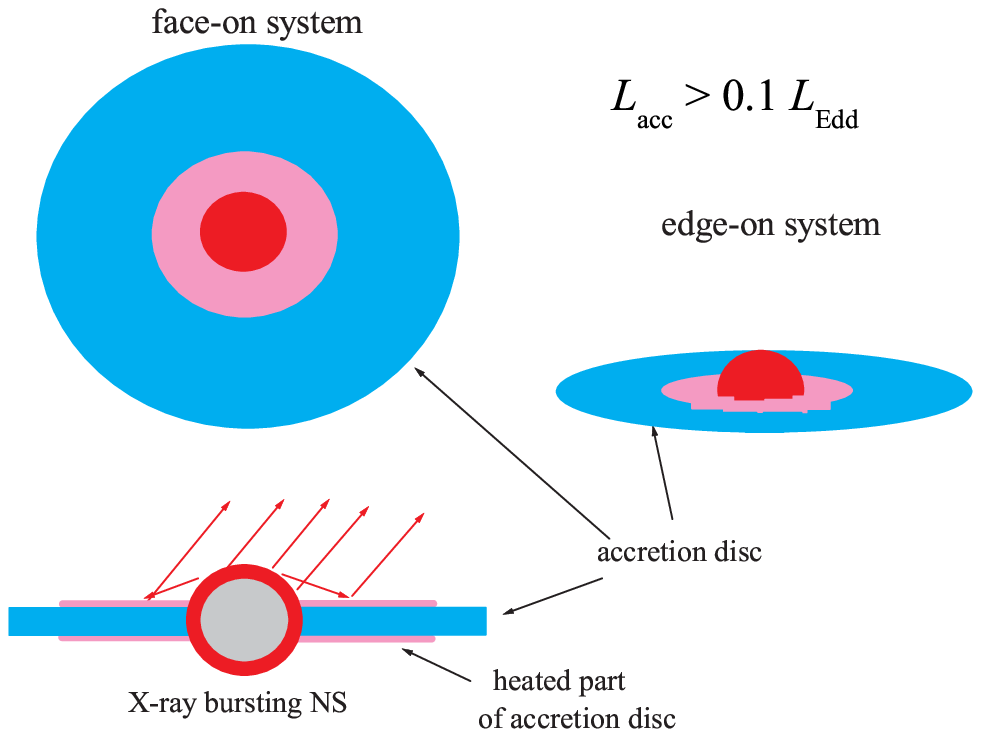}
  }
 \includegraphics{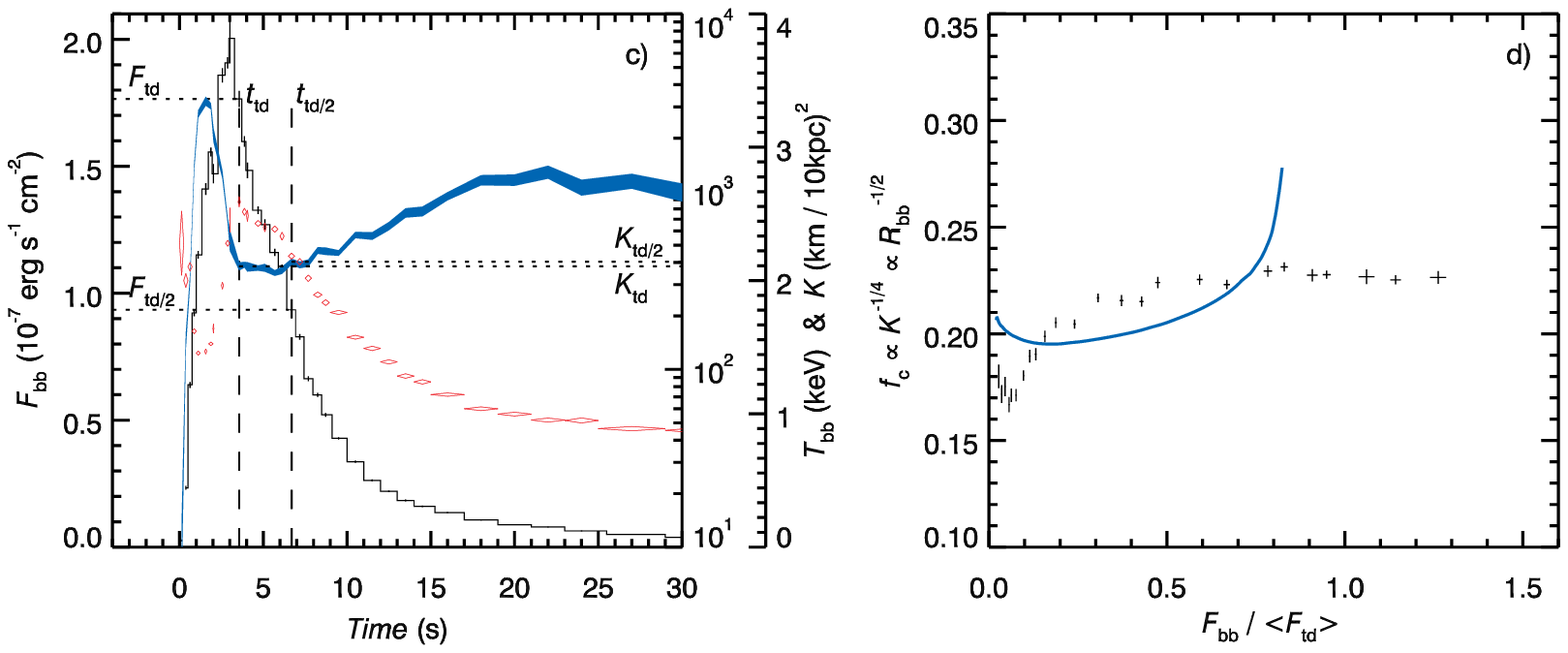}
\caption{Top panel: A scheme of the radial cross-section of the
  accretion structures around the NS in the soft persistent state of a LMXB
together with the projections on the sky for low (face-on) and high
(edge-on) inclination systems. 
 Bottom panel: The same as in fig.\,\ref{sv_f16}, but for the PRE
burst of 4U\,1608$-$52 in the soft persistent state. From ref. \cite{2014MNRAS.445.4218K}.
  }
\label{sv_f17}  
\end{center} 
\end{figure*}

The described method can be applied to the NS mass and radius measurements
if our assumption of a passively cooling ``isolated'' NS  
in empty space as a model of  an X-ray bursting NS during cooling phase
is correct. In reality, X-ray bursting NSs are 
accreting NSs with  the accretion discs around them. It is well known
that  LMXBs can have two qualitatively different 
persistent spectral states: the soft/high state and the hard/low state
(see fig.\,\ref{sv_f15}).  The luminosities of LMXBs in the soft
states are relatively high ($\sim 0.1\,L_{\rm Edd}$) and their spectra
can be composed of the soft ($kT < 1$\,keV) and the hard ($kT \sim
2-2.5$\,keV) components \cite{GRM03,RG06}. 
These components likely correspond to the optically thick 
accretion disc and to the optically thick boundary layer between the
accretion disc and the NS surface \cite{IS99,SulP06}. Low luminosity LMXBs ($L \sim 0.01 -
0.05\, L_{\rm Edd}$) show hard X-ray spectra described by 
emission models
of optically thin ($\tau_{\rm e} \sim 1$) hot ($kT \sim 20-30$\,keV)
thermal Comptonization. This means that the spectra are formed by 
multiple Compton scattering of the soft photons on the hot plasma
electrons \cite{ST80}.  
Therefore, optically thick discs do not exist in those states.
They are transformed to the geometrically thick optically thin hot
accretion flows, most probably due to the accretion disc evaporation 
\cite{1994A&A...288..175M}.  Intermediate spectral states with
 various contributions  of the optically thick
boundary layer,  the  
hot optically thin accretion flow as well as the standard 
accretion disc to the observed spectra are also observed.

It seems that the optically thin accretion flow could affect X-ray
burst flux less significantly compared to the optically thick disc. 
Moreover, the hot flow could be completely blown away by a PRE
burst. Therefore, we can expect that the best accordance of the real
X-ray  bursting NS with the ideal model occurs in PRE burst 
happening in the hard persistent state during the phase 
between the touchdown and a point when the accretion starts again.
The touchdown point ($t_{\rm td}$) is the moment of the PRE X-ray burst
with the maximum observed 
blackbody temperature $T_{\rm BB}$ and the minimum value of the
normalization $K_{\rm BB}$ \cite{GMH08}. It is commonly
accepted that
the NS photospheric size does not change after this moment, and the
corresponding bolometric flux $F_{\rm td}$ is close to the observed 
Eddington flux. Indeed, there is a number of PRE X-ray bursts in the hard
persistent states, which demonstrate    
a theoretically predicted dependences $K_{\rm BB}^{-1/4} - F_{\rm BB}$ in these
X-ray burst phases, see, e.g., the PRE bursts from 4U\,1608$-$52
(fig.\,\ref{sv_f16}) \cite{2014MNRAS.442.3777P,2014MNRAS.445.4218K,2015arXiv150906561N}. Therefore, we can use this kind of
PRE bursts to measure NS masses and radii.

\begin{figure}
\resizebox{0.45\textwidth}{!}{%
  \includegraphics{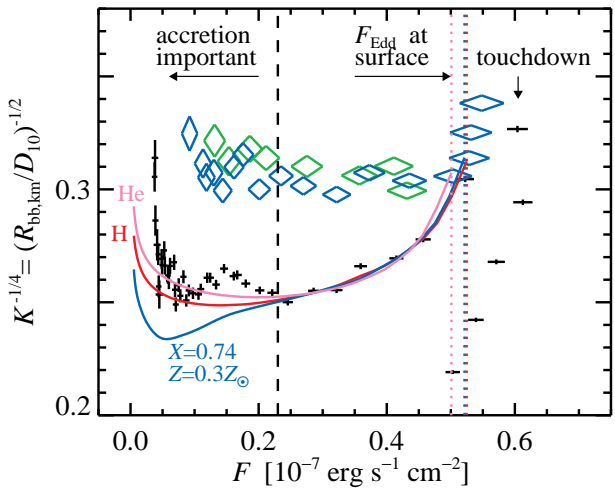}
}
\caption{A comparison of the X-ray burst data for 4U\,1724$-$307 with
  the theoretical models of the NS atmosphere. 
The crosses indicate the observed $K_{\rm BB}^{-1/4} - F_{\rm BB}$ dependence
for a PRE burst in the hard persistent state,  
while diamonds represent two PRE bursts in the soft persistent state. 
The solid curves correspond to the three best-fit theoretical models 
of various chemical compositions. 
  From ref. \cite{SPRW11}. 
   }
\label{sv_f18}   
\end{figure}

On the other hand, the optically thick accretion disc has to affect
the observed  
flux and spectrum of an X-ray burst significantly.  
It reflects the X-ray burst flux and 
blocks part of the bursting NS in high inclination systems
(fig.\,\ref{sv_f17}). This is a well known problem 
of radiation anisotropy of X-ray bursts \cite{LvPT93}
\be
      F_{E} = \xi_{\rm b}^{-1} \frac{L_{E, \infty}}{4\pi D^2}.
\ee 
The anisotropy coefficient  can be expressed with a simple formula \cite{1985MNRAS.217..291L}
\be \label{anis}
   \xi^{-1}_{\rm b} = \frac{1}{2} + \cos i,
\ee
where $i$ is the inclination angle, i.e. the angle between the
accretion disc axis and the line of sight. 
According to this approximation, $F_{\rm td}$ for  a PRE burst in
the soft persistent state can be  one and a half 
times higher than $F_{\rm td}$ for a PRE burst in the hard persistent
state in face-on systems ($\cos i \approx 1$). 
In fig.\,\ref{sv_f17} (bottom panel) an example of such a PRE burst in
4U\,1608$-$52 during the soft 
persistent state is shown.  
Indeed, its touchdown flux is about 1.5 times larger that the touchdown flux of other bursts during the hard state.
It is thus possible that 4U\,1608$-$52 is a face-on system. 
We also note that the dependence $K_{\rm BB}^{-1/4} - F_{\rm BB}$ is flat for
this burst and its behaviour 
is inconsistent  with the predictions  made on the base of eq.~(\ref{kfc}). 
It means that this equation cannot be applied to the PRE bursts in
the soft persistent state because of 
the accretion disc (whose effect is not completely understood) and the
influence of the boundary layer.  
Nevertheless, the PRE X-ray bursts occurring in the soft persistent states
with the flat $K_{\rm BB}^{-1/4} - F_{\rm BB}$ dependences have been widely used to determine NS parameters 
\cite{2010ApJ...719.1807G,2012ApJ...748....5O,2013ApJ...765L...1G}, 
leading obviously to strongly biased results.

LMXB 4U1724$-$307 is another example to study the difference between the soft and hard state bursts 
(see fig.\,\ref{sv_f18} and \cite{SPRW11}). 
The dependence $K_{\rm BB}^{-1/4} - F_{\rm  BB}$ for the PRE burst in the hard persistent state can be well
fitted with the  
theoretical $\fc - l$ dependence.  In the same figure, the dependences
$K_{\rm BB}^{-1/4} - F_{\rm BB}$ for two other PRE bursts in the soft
persistent states are shown. They are also flat like in 4U\,1608$-$52,
but the corresponding blackbody normalization $K_{\rm BB}$ is two times smaller than that for the
PRE in the hard persistent state. 
Unlike 4U\,1608$-$52, this source has   
similar maximum fluxes in the PRE bursts happening during the soft and in the hard
persistent states. We speculate that 4U\,1724$-$307 might be a highly inclined system,
and the accretion disc  blocks part of the NS in 
the soft persistent state, whereas the reflected flux has a
negligible contribution to the total luminosity. 
It is clear that the properties of the PRE bursts depend strongly 
on the spectral state of the persistent emission before the burst  \cite{2014MNRAS.442.3777P,2014MNRAS.445.4218K}.

\begin{figure}
\resizebox{0.45\textwidth}{!}{%
  \includegraphics{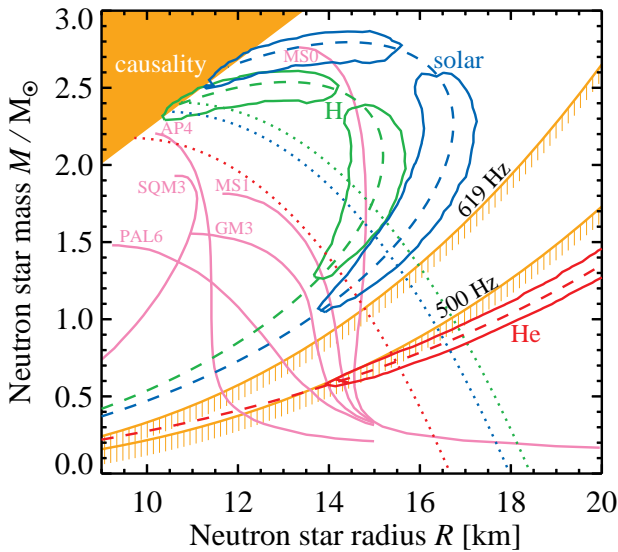}
}
\caption{ The mass--radius constraints from a hard-state PRE burst of
  4U\,1724$-$307 for three chemical compositions:  
green for pure hydrogen,  blue for the solar H/He ratio and
subsolar metal abundance $Z=0.3Z_{\odot}$, and red for pure helium 
assuming a flat distribution of the distance between 5.3 and 7.7 kpc
with  Gaussian tails of 1$\sigma$=0.6 kpc.  
The constraints are shown by contours (90\% confidence level). 
The mass-radius relations for several equations of state of neutron
and strange star matter are shown with solid pink curves.  
The brown solid curves in the lower right region correspond to the
mass-shedding limit and delineate the zone forbidden for the NS 
in 4U\,1724$-$307, if it had a rotational frequency of 500 or 619 Hz.
The dotted curves correspond to the best-fit parameter $A$ for the
distance to the source of 5.3 kpc.
  From ref. \cite{SPRW11}.
  }
\label{sv_f19}   
\end{figure}

\begin{figure}
\resizebox{0.45\textwidth}{!}{%
  \includegraphics{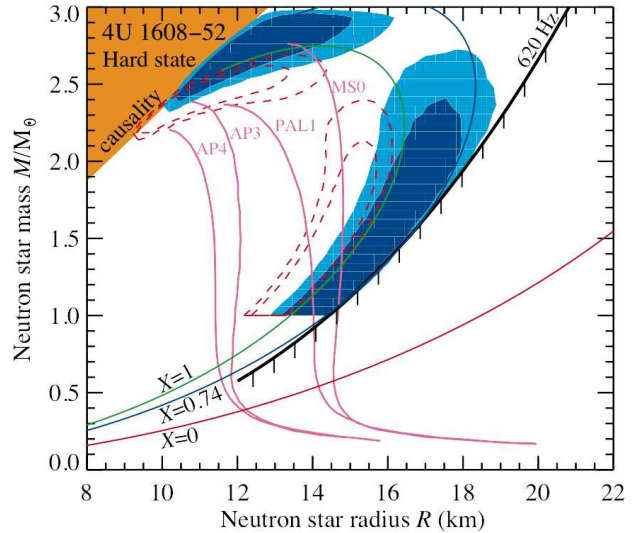}
}
\caption{The mass-radius constraints from the PRE bursts of
  4U\,1608$-$52 in the hard persistent state assuming $M >
  1.0\,M_\odot$, 
 distance 5.8$\pm$2.0 kpc, and a hydrogen rich atmosphere ($X>
0.74$). The dark and the light contours correspond to 68 and 90\% 
confidence levels. The red dashed contours give constraints 
with different data selection. 
The mass shedding limit for a spin frequency of
620\,Hz (measured in this source from coherent pulsations during 
some bursts) is marked by the black curve with downward ticks.  
The solid green, blue, and red curves marked with $X=1$, $X=0.74$, and
$X=0$ correspond to the best-fitting $T_{\rm Edd, \infty}$  
assuming  pure hydrogen, solar mix, and pure helium atmosphere
compositions. The NS mass-radius relations for several equation  
of states that have the maximum possible masses $> 2\,M_\odot$ are
shown with thin pink curves.   
  From ref. \cite{2014MNRAS.442.3777P}.
    }
\label{sv_f20}   
\end{figure}

We applied the cooling tail method to the PRE bursts of two LMXBs,
4U\,1724$-$307 \cite{SPRW11} and 4U\,1608$-$52  \cite{2014MNRAS.442.3777P},   
observed by the  \emph{RXTE} during their hard states. 
The obtained allowed regions in the $M-R$ plane for these NSs are
presented in figs.\,\ref{sv_f19} and \ref{sv_f20}.   
There are two main sources of uncertainties: the atmosphere chemical
composition and the distance.  
Hydrogen rich atmospheres give reasonable NS radii, 13$-$16 km, while 
helium rich atmospheres are excluded for both NSs because they give
too large ($> 20$\,km) radii for 1.5\,$M_{\odot}$ mass. 
 We note the solutions in figs.\,\ref{sv_f19} and \ref{sv_f20} are divided into two
parts and the region near the line $R= 4GM/c^2$ is  formally not allowed 
because of the degeneracy in the  transformation from  $(F_{\rm Edd},A)$ 
to $(M,R)$ \cite{2015arXiv150505156O}. 
The Bayesian analysis assuming flat priors in $M$ and $R$ instead of $F_{\rm Edd}$ and $A$ 
is free from this disadvantage, and the solutions close to  $R=4GM/c^2$ also become possible
\cite{2015arXiv150505156O,2015arXiv150906561N}.

Nevertheless, the obtained NS radii appear somewhat larger than 
the radius of the CCO in HESS\,1731$-$347 and the expected NS radii 
from  modern equation of states in NS cores \cite{2013ApJ...765L...5S}. 
A possible reason for this is a rapid rotation of the investigated NSs. 
The measured spin frequency of the NS in 4U\,1608$-$52 is 620\,Hz.
Such a rapid rotation distorts the NS shape making the equatorial
radius  larger than the polar one \cite{2014ApJ...791...78A}. 
As a result, an equivalent apparent radius of a rapidly rotating
NS becomes significantly larger than the radius of the non-rotating NS
of the same mass \cite{2015ApJ...799...22B}. 
Additionally, the combination of the shape distortion and the centrifugal force 
affects the surface gravity making it a function of the latitude, with the effective gravity being reduced at the equator. 
This in turn results in the latitude dependence of the  critical (Eddington) flux and 
reduction of the latitude-averaged critical luminosity compared to the 
Eddington limit for the non-rotating NS. Preliminary computations showed, that we can
expect that the radius of a non-rotating NS can be 1--2\,km smaller compared
to the measured radius of the NS in 4U\,1608$-$52.  
 We finally note that our most recent analysis of the carefully selected sample of 
hard-state bursts from 4U\,1702$-$429, SAX\,J1810.8$-$2609 and 4U\,1724$-$307 
produced  NS radii in the range 10.5--12.8 km \cite{2015arXiv150906561N} 
which is consistent with the estimates coming from the HESS\,1731$-$347 CCO.

\section{Summary}

We presented the basic ideas and methods allowing to measure  NS 
masses and radii from the analysis of their broad band X-ray
spectra. The model atmosphere spectra are the key
ingredient of the methods. We qualitatively described the main
features of the simplest NS model atmospheres together with the basic
assumptions used for NS model atmosphere computations. 
Our review is limited to non-magnetic NSs because the
construction of the magnetized NS model atmospheres 
is much more complicated and is associated with a large number of
uncertainties in the magnetized plasma opacities and with additional
geometry problems \cite{2011A&A...534A..74H}.    

There are two main effects making the model NS spectra harder than the
blackbody spectra with the same effective temperature:
(i) the decrease of the absorption opacity with the photon energy and
(ii) the increase of the atmospheric temperature with depth. 
Due of these atmosphere properties, the emergent photons preferentially escape
at higher energies where the atmosphere is 
more transparent. As a result,
the emitting NS sizes derived from blackbody  
fits are significantly smaller than the actual NS radii. Accurate
model atmospheres are necessary to take
these differences correctly into account. The additional opaque
spectral bands due to photoionization edges in the not fully
ionized atmospheres force emergent photons to escape at even harder
energies further increasing the deviation of the model atmosphere spectrum 
from the blackbody.

We also presented the previously published results of the NS mass and
radius measurements using the model NS spectra for two types of X-ray
sources: 
the  CCO in supernova remnant HESS\,1731$-$347 and X-ray bursting NSs in LMXBs, 4U\,1724$-$307 and 
 4U\,1608$-$52. The \emph{XMM-Newton} spectrum of the first 
 source was fitted with the carbon model atmosphere spectrum for the fixed
 distance of 3.2\,kpc with an additional limitation obtained from the 
 NS cooling theory for the assumed age of 10--40 kyr. The radius of 12--14
 km and the mass of 1.2--1.7\,$M_\odot$ were obtained for this NS. 
 The X-ray bursting NSs parameters were derived using a novel
 cooling tail method and the larger NS radii ($>$13 km for hydrogen
 rich  atmospheres and NS masses 1.0--2 $M_\odot$) were obtained. The NS in\\
 4U\,1608$-$52 rotates rapidly with the spin frequency of 620\,Hz. 
 We thus suggest an unaccounted  NS shape distortion to be the reason for
 this difference in the radii. The spin frequency of the NS in
 4U\,1724$-$307 is 
 unknown, but it is also likely to be high and the same explanation
 is valid for this NS as well.    
 On the other hand, our  recent analysis of  X-ray bursters from three LMXBs 4U\,1702--429, SAX\,J1810.8--2609 and 4U\,1724--307
 show the radii in the range 10.5--12.8 km. These results favour a reasonably stiff equation of state 
 of cold dense matter.

\begin{acknowledgement}

We thank J.~Kajava, J.~N\"attil\"a, G.~P\"uhlhofer, D.~Yakovlev, G.~G. Pavlov, A.~Santangelo and other co-authors
of the works used here for the successful collaborations. V.F.S. thanks DFG for financial support (grant WE 1312/48-1). 
J.P. acknowledges the Academy of Finland grant 268740. 

\end{acknowledgement}


\end{document}